\documentclass[a4paper]{article}
\usepackage[a4paper, total={6in, 9in}]{geometry}
\usepackage{floatrow}
\usepackage{hyperref}
\usepackage{filecontents}
\usepackage{subfigure}
\usepackage{bm}
\usepackage[font={scriptsize,bf}]{caption}
\usepackage{amsmath}
\usepackage{amsfonts}
\usepackage{amssymb}
\usepackage{graphicx}
\usepackage{bm}
\usepackage{multicol}
\usepackage{multirow}
\usepackage{epstopdf}
\usepackage{soul}
\usepackage{pifont}
\usepackage{lineno}
\usepackage[table]{xcolor}
\usepackage{color}
\usepackage[autostyle]{csquotes}
\usepackage{tikz,pgf}
\usepackage{collcell}

\usepackage{pgfplots}
\usepackage{wasysym}
\usepackage{lipsum}
\usepackage{algorithm} 
\usepackage{algpseudocode} 
\newlength\fheight
\newlength\fwidth
\newcommand{\cmmnt}[1]{\ignorespaces}
\usepackage{etoolbox}
\usepackage{setspace}
\usepackage{csquotes}

\title{A Numerical Investigation of the Aeroelastic Interaction between Transonic Buffet and Structural Nonlinearity}
\author{ \large Michael Candon$^1$\thanks{corresponding author, candon.michael@rmit.edu.au}, Vincenzo Muscarello$^1$, Pier Marzocca$^1$ and Oleg Levinski$^2$}

\date{
	\normalsize $^1$Department of Aerospace Engineering, RMIT University, Melbourne, AUS, 3000\\
	$^2$Defence Science and Technology Group, Fishermans Bend, AUS, 3207\\[2ex]%
}

\begin{document}
	
	\maketitle
	\begin{abstract}

Transonic shock buffet is a nonlinear, unsteady aerodynamic phenomenon characterized by self-sustained, periodic shock oscillations that can critically affect aircraft structural integrity. While the aerodynamic aspects of shock buffet have been widely studied, its interaction with nonlinear structural dynamics remains largely unexplored. This paper presents, for the first time, a numerical investigation of aeroelastic interactions arising from the coupling of shock buffet with a nonlinear structural model featuring pitch freeplay. Using unsteady Reynolds-Averaged Navier-Stokes (URANS) simulations coupled with a two-degree-of-freedom heave-pitch airfoil model, the study reveals that structural nonlinearity can induce aerodynamic lock-in to superharmonics of the heave natural frequency, resulting in 2:1 and 3:1 resonance mechanisms and large-amplitude heave limit cycles. These newly identified resonance behaviors expand the current understanding of transonic aeroelastic instabilities. The influence of key parameters such as structural-to-fluid mass ratio and structural damping on these phenomena is also systematically examined. This work introduces a novel class of aeroelastic lock-in mechanisms with significant implications for transonic flight dynamics and aircraft design.
        \end{abstract}

\section*{Nomenclature}
\begin{tabbing}
XXXXXXXXXX \= XX \= \kill
$b$ \>\> Semi-chord [m]\\
$c$ \>\> Chord [m]\\
$C_L$ \>\> Lift coefficient\\
$C_M$ \>\> Pitching moment coefficient\\
$C_p$ \>\> Pressure coefficient\\
$f$ \>\> Frequency [Hz]\\
$h$, $\dot{h}$, $\ddot{h}$ \>\> Heave displacement, velocity and acceleration [m],  [m/s],  [m/s$^2$]\\
$I_\alpha$ \>\> Airfoil moment of inertia [kg m$^2$]\\
$k$ \>\> Reduced frequency, $2\pi fc/u_\infty$\\
$k_{sb}$ \>\> Shock buffet reduced frequency, $\omega_{sb} c/u_\infty$\\
$\hat{k}$ \>\> Reduced frequency normalized by the shock buffet reduced frequency, $k/k_{sb}$\\
$\hat{k}_h$, $\hat{k}_\alpha$  \>\> Heave and pitch frequency ratio, $(\omega c/u_\infty)/k_{sb}$\\
$L$ \>\> Lift Force\\
$M_{c/4}$ \>\> Moment about the quarter-chord\\
$M_\infty$ \>\> Freestream Mach number\\
$m$ \>\> airfoil mass [kg]\\
$\bar{P_{xx}}$ \>\> Maximum normalized power spectral density\\
$t$ \>\> Time [s]\\
$u_\infty$ \>\> Freestream velocity [m/s]\\
$V^*$ \>\> Velocity index, $u_\infty/b\omega_\alpha \sqrt{\mu})$\\
$\alpha$, $\dot{\alpha}$, $\ddot{\alpha}$ \>\> Pitch rotation, rotational velocity and rotational acceleration [$^\circ$],  [$^\circ$/s],  [$^\circ$/s$^2$]\\
$\alpha_0$ \>\> Freestream angle-of-attack [$^\circ$]\\
$\alpha_s$ \>\> Freeplay angle [$^\circ$]\\
$\Delta C_L$ \>\> Lift coefficient peak-to-peak amplitude\\
$\Delta \tau$ \>\> Non-dimensional time-step\\
$\zeta$ \>\> Structural damping ratio\\
$\mu$ \>\> Structural-to-fluid mass ratio, $m/\pi\rho_\infty b^2$\\
$\rho_\infty$ \>\> Freestream fluid density [kg/m$^3$]\\
$\tau$ \>\> Non-dimensional time, $tu_\infty/c$\\
$\omega_h$, $\omega_\alpha$ \>\> Heave and pitch natural frequency [rad/s]\\
$\omega_{sb}$ \>\> Shock buffet frequency [rad/s]\\
\end{tabbing}

\section{Introduction}

In recent decades, increased access to high-performance computing and cutting-edge numerical methods have allowed researchers to tackle novel and increasingly challenging problems across the aerospace sciences. Within the realm of unsteady aerodynamics and aeroelasticity, there has been a major interest in nonlinear aeroelastic systems of increasing complexity, including turbulent nonlinear transonic flow phenomena, nonlinear structural mechanisms, and high-speed flows~\cite{riso24}. One area that has yet to receive significant attention is the study of transonic aeroelastic systems that contain both nonlinear aerodynamic and structural mechanisms. A few authors have addressed a simplified version of this problem, $i.e.$, when high amplitude limit cycle oscillations induced by structural nonlinearity introduce nonlinear time-invariant aerodynamic forcing~\cite{he17, he20, candon24b, candon24c}. However, more complex nonlinear aerodynamic mechanisms associated with global flow instabilities in the transonic regime have not been studied when coupled with a nonlinear structural model.  
        
        Transonic shock buffet is a global flow instability that occurs within a narrow range of the transonic regime due to complex shockwave boundary layer interactions. Shock buffet is a highly nonlinear phenomenon characterized by large amplitude self-sustained periodic shock oscillations that occur in the absence of structural motion. An excellent review of the authoritative works is given by Giannelis~\textit{et al.}~\cite{giannelis16}. Shock buffet has been studied experimentally on rigid airfoil models~\cite{mcdevitt85, jacquin09} and wings~\cite{koike16,sugioka21}. Simulation of this complex viscous unsteady aerodynamic phenomenon requires computational fluid dynamics codes. The majority of works consider Unsteady Reynolds Averaged Navier-Stokes (URANS) codes~\cite{ionovich12, giannelis18} while, more recently, increased access to high-performance computing has seen an increase in the use of scale resolving codes with an up-to-date review provided by Lusher \textit{et al.}~\cite{lusher24}. 

        Some authors have also delved into elastic problems where the lock-in phenomenon can be observed, $i.e.$, where the shock buffet oscillations exhibit lock-in to the structural natural frequency - causing significant amplification of the limit cycle oscillations (LCOs). Before reviewing the authoritative works on the topic, the frequency ratio must be defined - referring to the ratio of the structural reduced frequency, $k_n$, to the shock buffet reduced frequency, $k_{sb}$, $\hat{k_n} = k_n/k_{sb}$. Raveh and Dowell~\cite{raveh14b} provide one of the pioneering works on the topic, showing that for a single-degree-of-freedom (s-DOF) pitching airfoil, when the frequency ratio is below 1, the system responds as a forced harmonic oscillator with a relatively benign LCO oscillating at $k_{sb}$. They go on to show that when the frequency ratio is above 1, lock-in can occur where the system response shifts to $k_n$ and the LCO amplitude grows by more than one order of magnitude. In the same paper, Raveh and Dowell~\cite{raveh14b} extend the study to a two-degree-of-freedom (2-DOF) airfoil, where the heave mode has a frequency ratio below 1 and the pitch mode above 1, showing that the pitch mode exhibits lock-in and significant amplification while the heave mode does not. Another important finding is that the lock-in mechanism and associated LCO amplification have relatively low sensitivity to the structural-to-fluid mass ratio and high sensitivity to structural damping. Giannelis~\textit{et al.}~\cite{giannelis16} conduct a sweep of the frequency ratio for an elastic s-DOF pitching airfoil, demonstrating that lock-in is encountered at a frequency ratio of 1 and, for the undamped system, continues up to a frequency ratio of approximately 1.8. They also show that the system is very sensitive to structural damping, predominantly influencing the point of lock-off. It is shown that lock-in and the associated LCO amplitude is less sensitive to the structural-to-fluid mass ratio which primarily acts by slowing the growth of the oscillations. Qualitatively similar findings are provided by Quan~\textit{et al.}~\cite{quan15} who show that both structural damping and the mass ratio influence the size of the lock-in region, specifically, as structural damping and mass ratio increase, the point of lock-off decreases. Gao \textit{et al.}~\cite{gao17} propose a linear reduced order modelling approach which identifies the underlying mechanism of lock-in as a coupled mode flutter, $i.e.$, a coupling of a structural mode with a fluid mode. A review of transonic buffet in an aeroelastic setting is provided in by Gao~\textit{et al.}~\cite{gao20}. 
        
        Freeplay is a well-known structural nonlinearity that refers to the loosening of aircraft mechanical components over time due to excessive loading. Freeplay is characterized by a zero-stiffness deadzone in the stiffness distribution of a hinge that can be represented numerically as a bilinear spring~\cite{conner97}. The reason that freeplay has been the subject of many academic studies in the last few decades is that it can reduce the flutter boundary predicted by linear theory, by way of dangerous subcritical aeroelastic instabilities, and even introduce high-amplitude LCO~\cite{lee99}. Although freeplay is typically related to control surface actuation~\cite{kousen94, kim00, dowell03, candon19a}, it has also been problematic for other dynamics affected components, such as wing-fold hinges~\cite{lee06} or store linkages~\cite{tang06a}.

        A natural and, to the authors’ knowledge, previously unexplored progression in aeroelastic research involves the study of shock buffet on a nonlinearly suspended airfoil—specifically, one that incorporates structural nonlinearities such as cubic hardening/softening, bilinear stiffness (freeplay), or nonlinear damping. This combined nonlinear aero-structural problem is not only compelling from a nonlinear dynamical systems perspective but also highly relevant to aircraft certification and structural sustainment. Understanding the two-way interaction with control surface freeplay(s) is critical for the following reasons:

        \begin{itemize}
            \item The effect of freeplay on the lock-in mechanism is unknown.
            \item  Freeplay growth is significantly impacted by large amplitude cyclic loading, such as the loads experienced in buffeting flow.
            \item Neglecting the interaction between transonic shock buffet and freeplay could lead to an underestimation of the airframe buffet loading, which might significantly impact fatigue life predictions~\cite{levinski12, levinski20}.
        \end{itemize}
        
        Of particular interest is the first item; the influence of freeplay on the lock-in mechanism. A well-known artifact of structural nonlinearity is the redistribution of vibrational energy from the fundamental modes, $\omega_n$, to their integer multiples $p\omega_n, \, p = 2, 3, 4, ...$ (superharmonics), and fractional multiples $\omega_n/p, \, p = 2, 3, 4, ...$ (subharmonics). An undesirable consequence of these additional vibration modes in the system is nonlinear resonance. Specifically, if the frequency of a periodic external forcing function coincides with a super- or subharmonic, it can cause large amplitude oscillations that would not have otherwise occurred in the absence of the structural nonlinearity. 

        In this paper, a two-degree-of-freedom (2-DOF) heave-pitch aeroelastic airfoil of the NACA0012 profile is used to conduct a numerical investigation of the buffet-freeplay interaction. The unsteady aerodynamic forces are computed using a URANS code with embedded nonlinear structural equations-of-motion. The objective is to understand if dangerous large amplitude aeroelastic instabilities can occur at frequency ratios below the typically cited range ($\hat{k_n} << 1$) by way of lock-in to superharmonics of the systems natural frequencies. It is shown that, with a sufficiently large pitch freeplay, the aerodynamic forces can lock-in to superharmonics of the heave mode, causing resonance at ratios 2:1 and 3:1. These subharmonic resonances and other rich nonlinear dynamical features are studied in detail, along with the influence of structural-to-fluid mass ratio and structural damping. 

        The remainder of the paper is organized as follows; in Section II the aeroelastic equations-of-motion are formulated and the computational fluid dynamics model is described. In Section III, validation and mesh refinement are performed. The results are presented and discussed in Section IV, and Section V provides a final discussion and concluding remarks. 

    \section{Computational Framework}
	
        This study considers a widely used shock buffet benchmark based on a two-dimensional NACA0012 airfoil model. The experimental campaign was carried out by Mcdevitt and Okuno~\cite{mcdevitt85} in the NASA Ames Research Center high-Reynolds number facility. The tests were performed at Mach numbers ranging from approximately 0.7 to 0.8, identifying the buffet onset angle, and at Reynolds numbers from $1\times10^6$ to $14\times10^6$. The experiment was designed specifically to obtain two-dimensional airfoil data with minimum wall interference. 
	

    \subsection{Computational Fluid Dynamics Model}
    The general purpose finite volume code ANSYS Fluent 2024 R1~\cite{ansys} is used with a coupled pressure-based implicit solver and Rhie-Chow distance-based flux interpolation. Second-order upwind differencing is used for the advective fluxes and central-differencing for the diffusive terms. A dual time-stepping scheme is employed with second-order implicit temporal discretization and a non-dimensional time-step of $\Delta \tau = 8\times 10^{-3}$. The Spalart-Allmaras (SA) turbulence model~\cite{spalart92} is used for all simulations with the Spalart-Shur curvature correction (CC)~\cite{spalart97}, which significantly improves the prediction of shock buffet. 

\begin{figure}[h]
        \centering
        \includegraphics[width=0.5\textwidth]{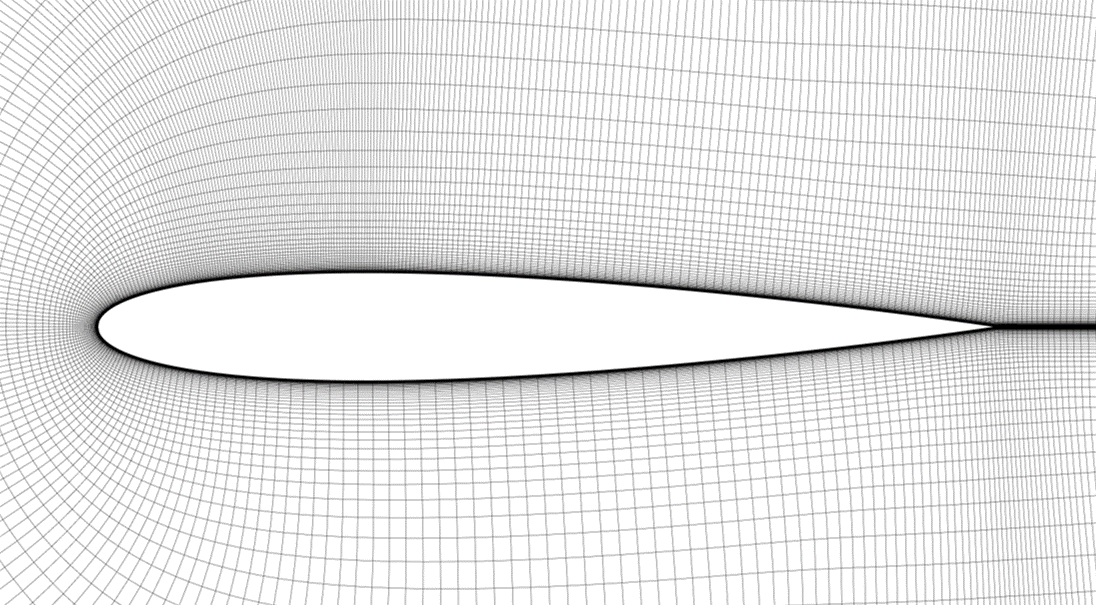}
    \caption{Computational grid c}
    \label{fig:mesh}
    \end{figure}

    Four C-grid topologies are constructed using Pointwise to assess mesh independence, with grid C presented in Fig.~\ref{fig:mesh}. The grid extends 80 chord lengths in the upstream and normal directions, and 100 chord lengths in the downstream direction. Mesh refinement follows the strategy presented by Ionovich and Raveh~\cite{ionovich12}, focused primarily on region of the airfoil across which the shock traverses (10\% - 30\% of the chord). Mesh statistics of the four grids are provided in Table~\ref{tab:mesh}. The first cell height is $3\times10^{-6}c$ for all grids, producing a maximum $y^+$ less than unity.

    \begin{table}[h]
    \centering
    \begin{tabular}{ccccc}
        \hline
          \textbf{metric} &\textbf{A} &\textbf{B} &\textbf{C} &\textbf{D} \\
        \hline
        size & 356$\times$95 & 393$\times$107 & 444$\times$107& 541$\times$123 \\
        growth rate & 1.2 &  1.15 & 1.15 & 1.13 \\
        shock resolution $c$ & 0.009 &  0.0045 & 0.003& 0.0015 \\
        \hline
    \end{tabular}
    \caption{Grid refinement statistics}
    \label{tab:mesh}
    \end{table}

\subsection{Nonlinear Aeroelastic Equations-of-Motion}
As per the work of Raveh and Dowell~\cite{raveh14b}, this research considers the uncoupled 2-DOF aeroelastic equations-of-motion with the addition of freeplay, given by:
        \begin{subequations}
            \begin{align}
            m(\ddot{h} + 2\zeta_h \omega_h \dot{h} + {\omega_h}^2h) = L,\\
            I_\alpha(\ddot{\alpha} + f(\alpha,\dot{\alpha})) = M_{c/4}
            \end{align}
        \end{subequations}

 \noindent where $h, \dot{h}, \ddot{h}$ are the heave displacement, first and second time derivatives, $\omega_h$ is the heave natural frequency, $\zeta_h$ is the heave mode damping ratio, $L$ is the sectional lift, $I_\alpha = \mu \pi \rho_\infty b^4 r_\alpha^2$ is the moment of inertia, $\rho_\infty$ is the freestream fluid density, $b = 0.5$m is the semi-chord, $r_\alpha^2 = 1$ is the radius of gyration, $\mu = 75$ is the nominal structural-to-fluid mass ratio which is based on an airfoil mass of $m = 3$kg, and $M_{c/4}$ is the sectional moment taken about the quarter chord. Under the assumption of zero structural damping within the freeplay dead-zone, the freeplay nonlinearity is described by a piece-wise function $f(\alpha,\dot{\alpha})$:

 \begin{equation}
f(\alpha,\dot{\alpha}) = \left\{ \begin{array}{lllll} 2\zeta_\alpha \omega_\alpha \dot{\alpha} + {\omega_\alpha}^2(\alpha-\alpha_s)  &&&& \alpha > \alpha_s \\ 0  &&\text{if}&& -\alpha_s < \alpha < \alpha_s \\ 2\zeta_\alpha \omega_\alpha \dot{\alpha} + {\omega_\alpha}^2(\alpha+\alpha_s) &&&& \alpha < \alpha_s \end{array} \right.
\end{equation}

\noindent where $\alpha, \dot{\alpha}, \ddot{\alpha}$ are the pitch rotation, first and second time derivatives, $\omega_\alpha$ is the pitch natural frequency, $\zeta_\alpha$ is the pitch mode damping ratio and $\alpha_s$ is the half rotational freeplay gap. The structural equations-of-motion are embedded in ANSYS Fluent via User Defined Function, with a fourth-order Runge-Kutta scheme used to converge the structural motion. A dynamic mesh model is used to capture the induced momentum due to the motion of the wing based on diffusive smoothing, which preserves mesh quality close to the boundary whilst absorbing motion in the farfield. 

    \section{Mesh Independence and Experimental Validation}

    Validation of the SA model is conducted for both steady (pre-buffet) and unsteady (buffet) cases. The operating conditions of the steady case are $M_\infty = 0.72$, $\alpha_0 = 4^\circ$ and $Re \approx 1\times10^7$, which the experiment showed was below the buffet onset angle. The operating conditions of the unsteady case are $M_\infty = 0.72$, $\alpha_0 = 6^\circ$ and $Re \approx 1\times10^7$, where significant shockwave motion is reported in the experiment~\cite{mcdevitt85}. 

\begin{figure}[h]
		\centering
			\includegraphics[width=1\textwidth]{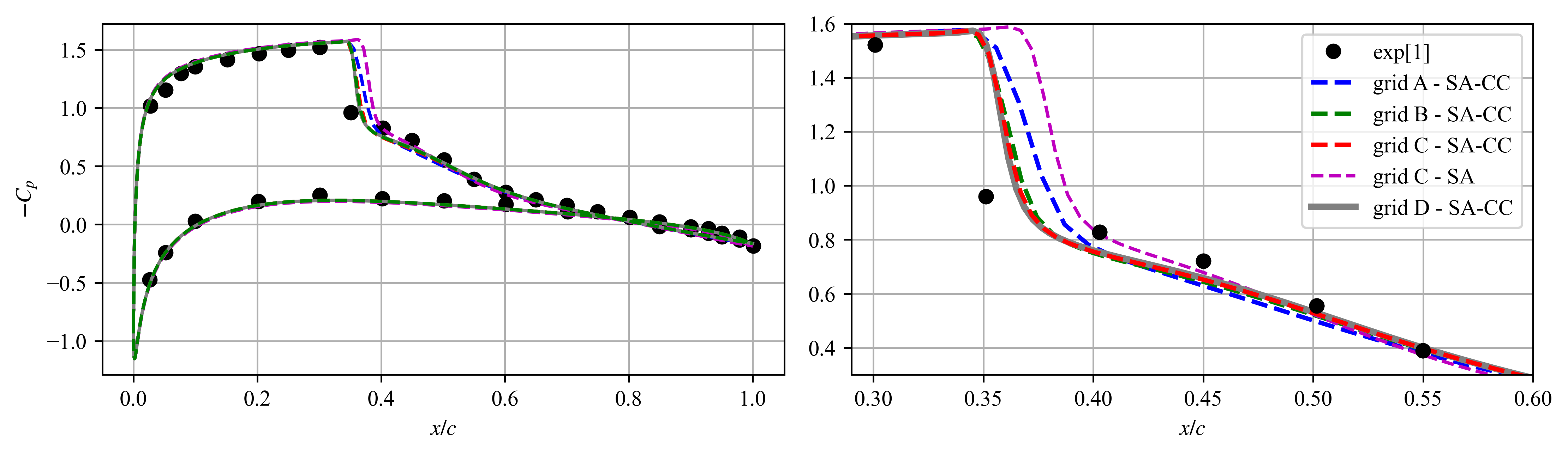}
		\caption{Grid refinement for pressure coefficient distribution at the pre-buffet condition $M_\infty = 0.72$, $\alpha_0 = 4^\circ$ and $Re \approx 1\times10^7$}
		\label{fig:mesh_independenceA}
	\end{figure}

    Figure~\ref{fig:mesh_independenceA} illustrates the grid sensitivity to capturing the pressure coefficient on the airfoil surface for the steady case. All grids demonstrate that the predicted shock location is farther downstream than the experiment, which is consistent with the results of Ionovic and Raveh~\cite{ionovich12}. In terms of the numerical results, there is minor discrepancy in the result of the coarse grid (A), while convergence between the fine (C) and super-fine (D) grids is evident. The result of the fine grid without curvature correction is also given. where it can be seen that the error in prediction of shock location worsens. For the unsteady buffeting case, the buffet statistics are given in Table~\ref{tab:mesh_independence}. Convergence of $\Delta C_L$ is also observed for the fine grid, and the buffet reduced frequency demonstrates a discrepancy of 1.29\%. The error compared to the experiment for the fine grid is a 6.1\% under-prediction. The buffet onset angle is over-predicted by 5.12\%. With these findings, the fine grid is used for the remainder of the paper.

    \begin{table}[h]
    \caption{Grid refinement and validation at the buffet condition $M_\infty = 0.72$, $\alpha_0 = 6^\circ$ and $Re \approx 1\times10^7$}
        \label{tab:mesh_independence}
    \begin{tabular}{ccccccc}
        \hline
         & exp.~\cite{mcdevitt85} & A & B & C & D \\
        \hline
        $\Delta C_L$ & -& 0.4921 & 0.4886 & 0.4833 &0.4852  \\
        buffet onset, $\alpha_{sb}$ [$\circ$] & 4.1 & - & - & 4.31 & - \\
        buffet reduced frequency, $k_{sb}$ & 0.55& 0.5129 & 0.5138 & 0.5167 & 0.5195 \\
        \hline
    \end{tabular}

    \end{table}

\section{Results}
\label{sec:res}

All results presented in this section consider the freestream conditions of $M_\infty = 0.72$, $\alpha_0 = 6^\circ$ and $Re \approx 1\times10^7$. The investigation considers a variable heave frequency ratio of $\hat{k}_h = 0.4 - 0.555$ and a pitch frequency ratio of $\hat{k}_\alpha = 0.755$ with various pitch freeplay angles. The aeroelastic simulations are started after the buffet oscillations on the rigid airfoil have reached a bounded limit cycle, which are monitored via the integrated aerodynamic quantities. The maximum number of physical time-steps that are simulated to assess if lock-in occurs is $1\times10^6$. 
    
\subsection{Flutter}
To verify that any LCO observed is due to the unsteady shockwave motion only and not the result of a post-flutter instability, it is necessary to assess the flutter characteristics of the system. This is a challenging prospect in a buffeting condition~\cite{gao17} considering that linear flutter analysis requires a stable aerodynamic condition. Raveh and Dowell~\cite{raveh14} approximate the flutter speed using Theodorsen aerodynamics, while Gao~\textit{et al.}~\cite{gao17} use a flow stabilization approach. 

\begin{figure}[h]
    \centering
        \includegraphics[width=0.5\textwidth]{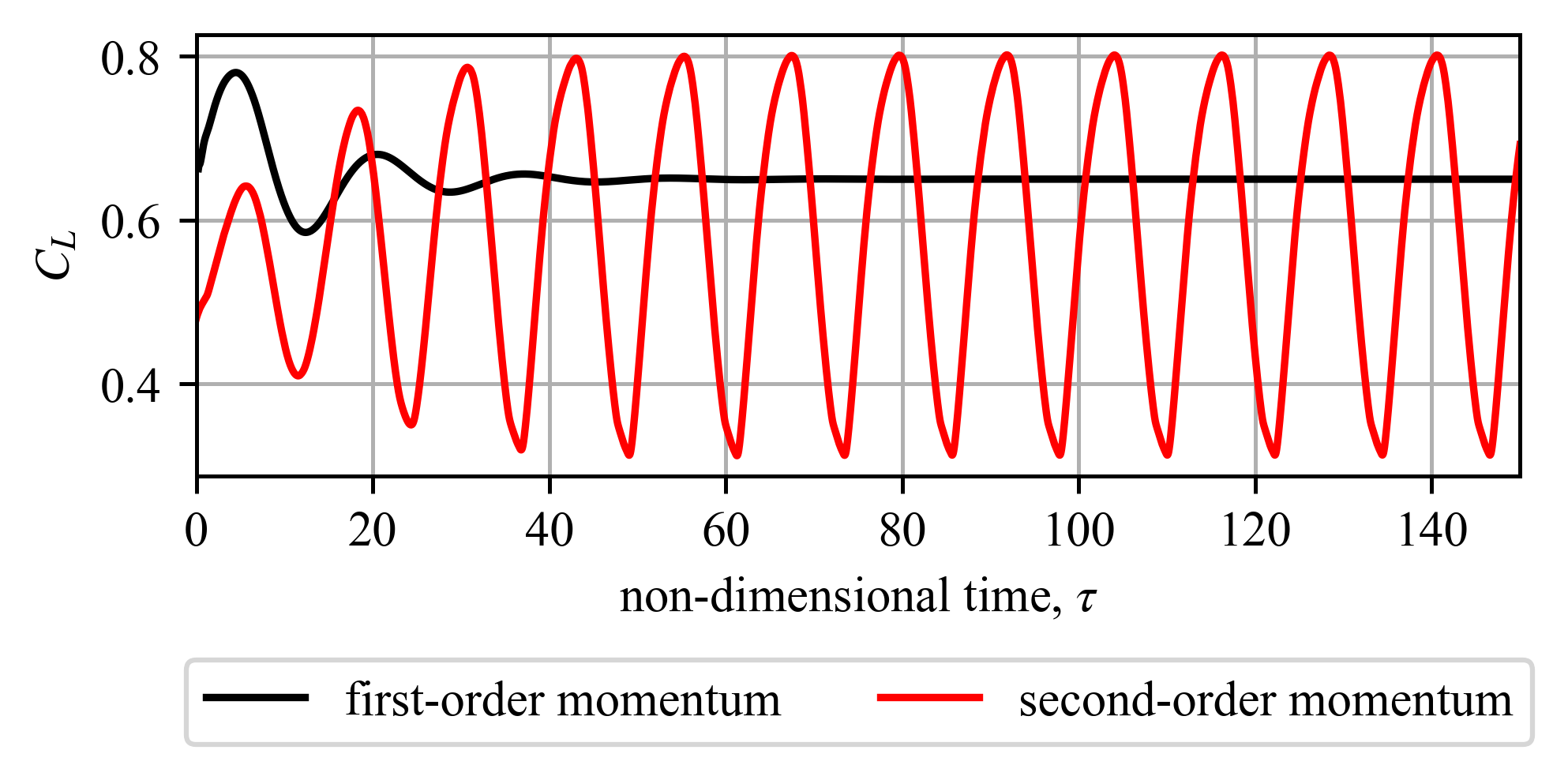}
    \caption{Lift coefficient comparing first-order and second-order discretization of the momentum equation at the nominal operating condition}
    \label{fig:fo_vs_so}
\end{figure}

\begin{figure}[h!]
    \centering
        \subfigure[first-order momentum (steady)]{\label{mesh1}
        \includegraphics[width=0.45\textwidth]{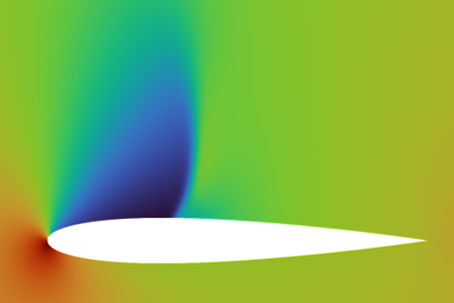}}
        \subfigure[second-order momentum (rms)]{\label{mesh1}
        \includegraphics[width=0.45\textwidth]{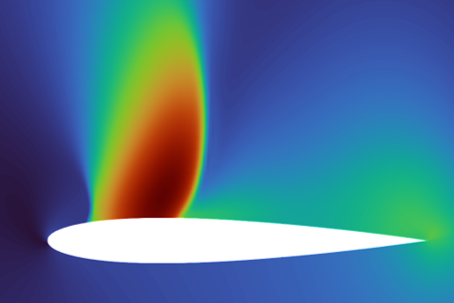}}
    \caption{Pressure coefficient contours comparing first-order and second-order discretization of the momentum equation at the nominal operating condition.}
    \label{fig:pres_coef}
\end{figure}

In this work, the linear flutter characteristics of the system are approximated by first reducing the spatial discretization of the momentum equation to first-order accuracy - artificially removing the shock buffet oscillations through increased numerical diffusion as can be observed in Fig.~\ref{fig:fo_vs_so}. Then a linearized frequency domain flutter solution is computed using CFD-based indicial aerodynamics. The instantaneous pressure coefficient of the stable first-order solution, and the RMS of the second-order solution, are presented in Fig.~\ref{fig:pres_coef}. It can be seen that the stable first-order solution approximates the shock location towards the maximum downstream location seen in the buffeting flow.  This approach can give an approximate estimation of the linear flutter characteristics, which is deemed to be sufficient for the purposes of this work. 

\begin{figure}[h]
    \centering
        \includegraphics[width=0.5\textwidth]{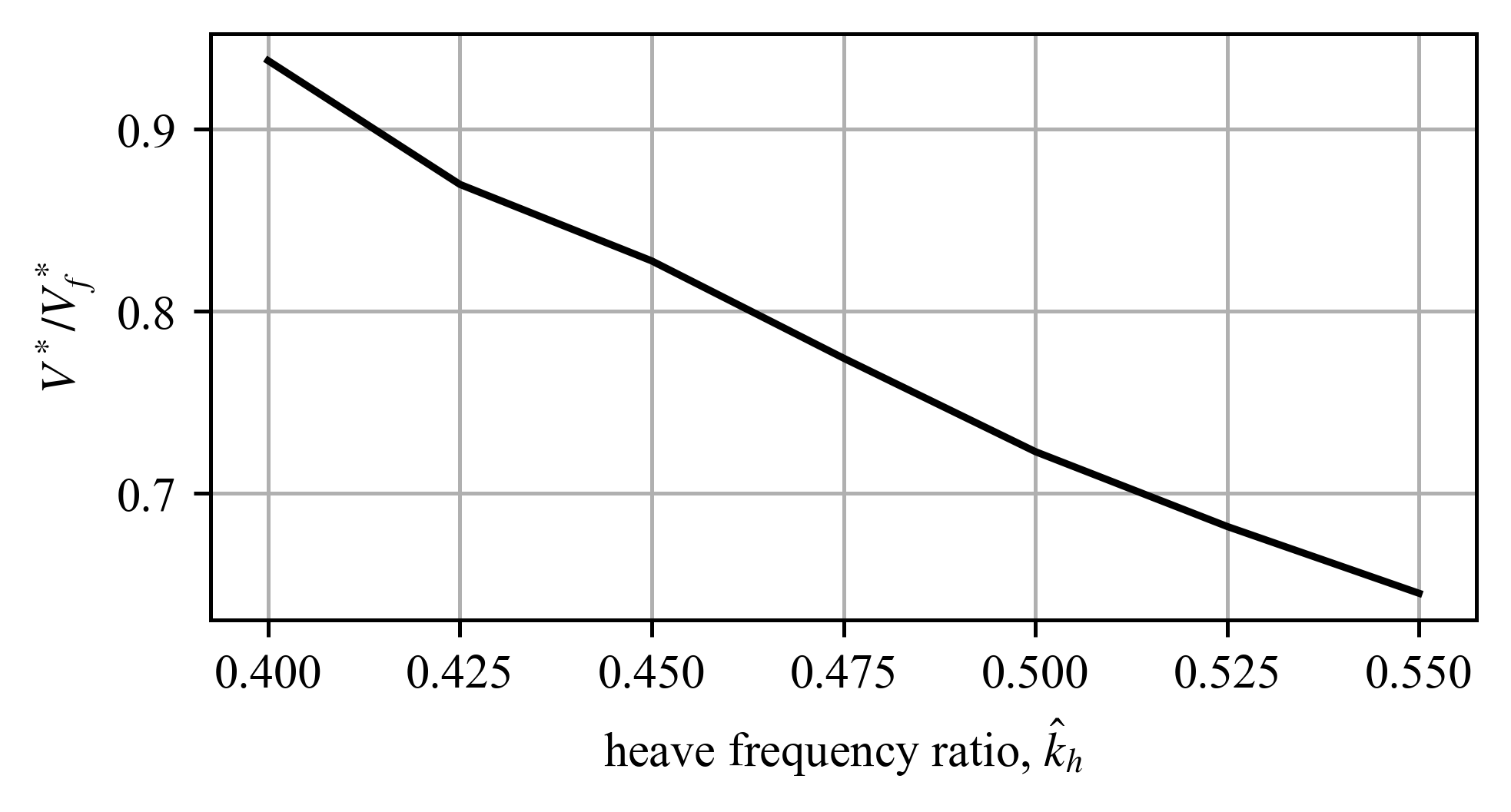}
    \caption{Flutter speed ratio at the nominal operating condition for varying heave natural frequencies.}
    \label{fig:flutter_rfa}
\end{figure}

The linearized frequency domain flutter solutions consider a least-squares rational function approximation (RFA)~\cite{roger77} of the indicial aerodynamic forces that are computed via the CFD code. More details on the RFA approach are provided in work by the authors~\cite{candon19phd}. The frequency domain computation of the ratio of the operational velocity index to the velocity at flutter, $V^*/V^*_f$, for the heave natural frequencies of interest, are presented in Fig.~\ref{fig:flutter_rfa}. It can be seen that the system is below the linear flutter speed in all cases. As the heave natural frequency reduces, the operating condition approaches the linear flutter speed. At the lowest heave natural frequency, where the heave frequency ratio is $\hat{k}_h = 0.4$, the system is operating at 96\% of the flutter speed.  

\begin{figure}[h]
    \centering
        \includegraphics[width=1\textwidth]{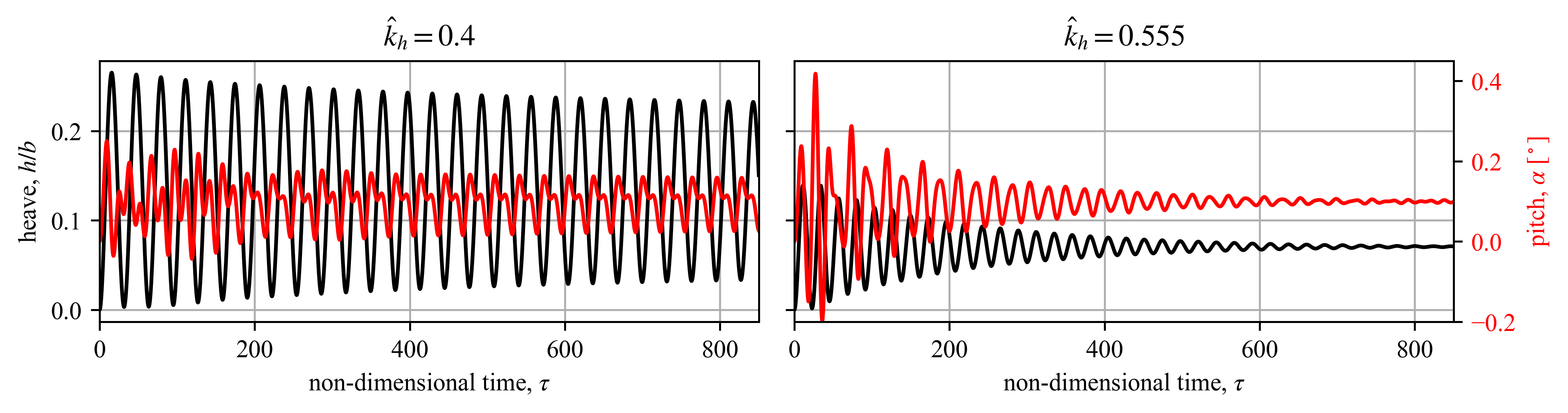}
    \caption{Time-domain aeroelastic response with first-order discretization of the momentum equation.}
    \label{fig:flutter_cfd}
\end{figure}

Given the relatively high angle-of-attack (AOA), there is also risk of a single-degree-of-freedom (s-DOF) flutter mechanism as indicated in an experimental campaign of a NACA0012 semi-span model conducted in the NASA Transonic Dynamic Tunnel (TDT) by Rivera \textit{et al.}~\cite{rivera92}. For a comprehensive discussion on the topic of s-DOF flutter see the technical note of Dowell~\cite{dowell24}. Recent work in s-DOF flutter modeling has shown that linear flutter tools may not always be able to predict it, particularly pertinent to the transonic regime~\cite{stanford24,candon25a}. Therefore, to ensure that a s-DOF aeroelastic instability is not encountered, the aeroelastic response at the minimum and maximum heave natural frequency is computed using the time-marching CFD solver as presented in Fig.~\ref{fig:flutter_cfd}. It can be seen that at $\hat{k}_h = 0.4$ the response mildly damped and at $\hat{k}_h = 0.55$ is highly damped, in good agreement with the frequency domain predictions and confirming that the system is not encountering a s-DOF flutter mechanism. 

\subsection{Heave Frequency Ratio Sweep}
\label{sec:frat_sweep}
In this section, a sweep of the heave natural frequency, $0.4 \leq \hat{k}_h \leq 0.555$ is conducted, with varying degrees of freeplay, which unveils a range of novel findings on the topic of aeroelastic shock buffet interactions. 

\begin{figure}[h]
    \centering
        \includegraphics[width=1\textwidth]{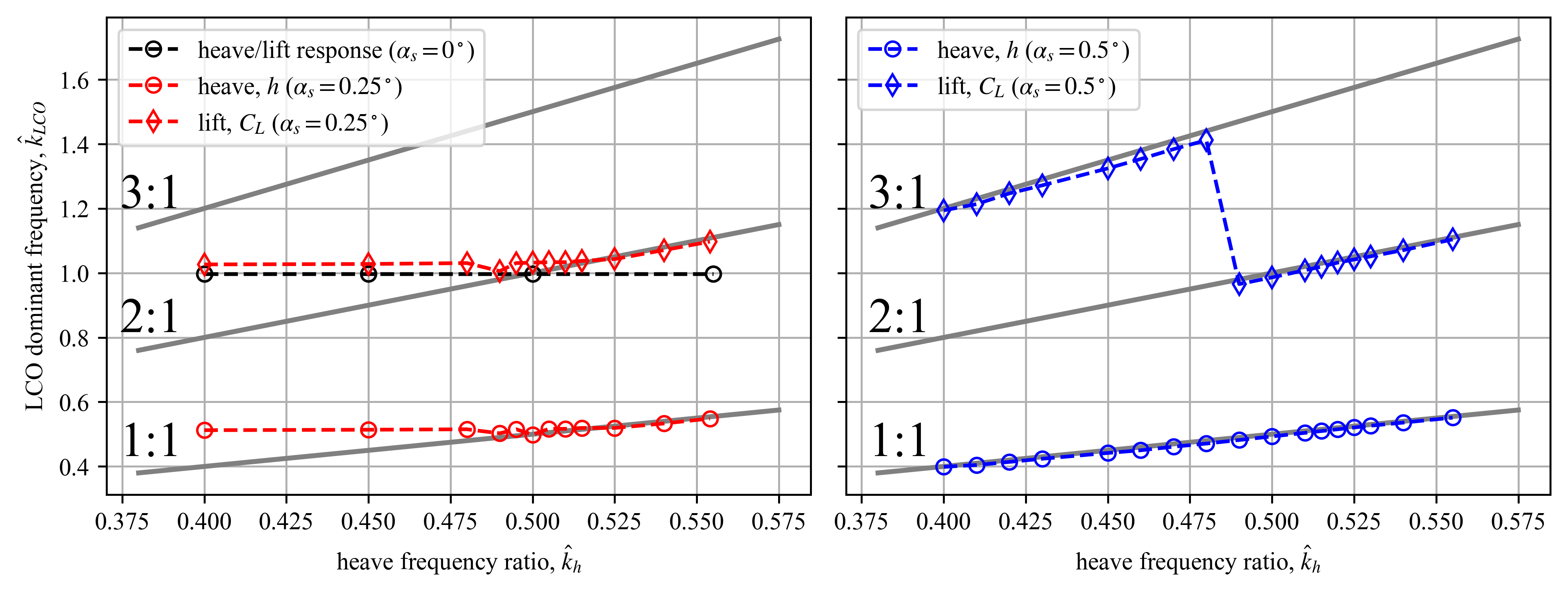}
    \caption{Dominant response frequencies with $\mu = 75$, $\hat{k}_h = 0.4-0.555$, $\hat{k}_\alpha = 0.755$}
    \label{fig:freq_rat_sweep_freq}
\end{figure}

\begin{figure}[h]
    \centering
        \includegraphics[width=1\textwidth]{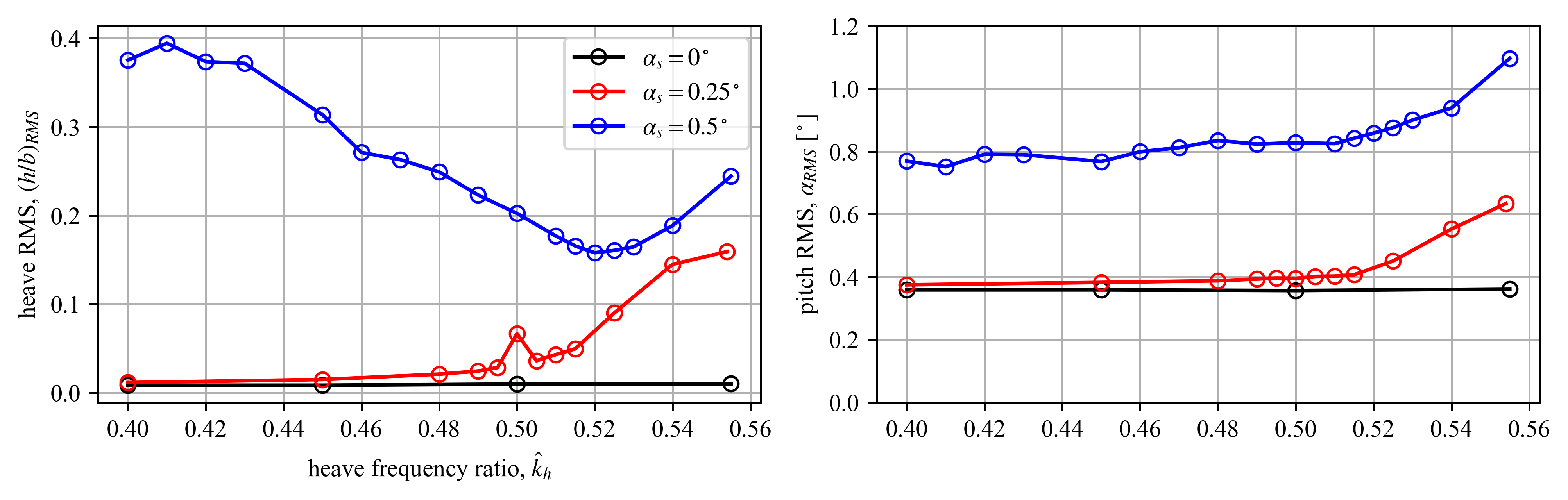}
    \caption{Response RMS with $\mu = 75$, $\hat{k}_h = 0.4-0.555$, $\hat{k}_\alpha = 0.755$}
    \label{fig:freq_rat_sweep_rms}
\end{figure}

Figure~\ref{fig:freq_rat_sweep_freq} presents the dominant frequency of the aerodynamic and structural stable LCO, also normalized by the shock buffet frequency, $\hat{k}_{LCO}$, as a function of the installed heave frequency ratio, $\hat{k}_h$. The dominant frequency here is defined by the frequency with maximum power as calculated via the power spectral density. Given that the stable heave and pitch limit cycles have equivalent dominant frequencies, as do the stable lift and moment limit cycles, only heave and lift are presented here. The corresponding RMS of the structural responses are presented in Fig.~\ref{fig:freq_rat_sweep_rms}. 

Initially, looking at the case without freeplay, it can be seen that both the aerodynamic and structural LCOs are oscillating at the shock buffet frequency, for all natural frequencies. For these cases, the heave and pitch RMS LCO amplitude is small and does not see any notable amplification with the variation of the heave frequency ratio. This behavior is aligned with the findings of other authors~\cite{raveh14, giannelis16}, specifically when both modes are characterized by a frequency ratio well below 1, no form of lock-in occurs and the system behaves as a forced harmonic oscillator.  

Next, looking at the case with freeplay $\alpha_s = 0.25^\circ$, at lower heave frequency ratios $\hat{k}_h<0.5$, the lift forces have a consistent dominant frequency that is slightly above the shock buffet frequency. The structural response is at exactly half of this. No notable amplification of the heave or pitch LCO is encountered for these cases. At a frequency ratio of $\hat{k}_h = 0.5$ a structural resonance is encountered, where the structure oscillates at the heave natural frequency, leading to some amplification of the heave mode response. However, at this condition, the aerodynamics does not lock-in given the slight offset of the dominant aerodynamic forcing. Lock-in occurs as the heave frequency ratio moves beyond $\hat{k}_h = 0.515$ (half the aerodynamic forcing frequency), the aerodynamics locks in to the heave natural frequency superharmonic $2k_h$, and a 2:1 resonance is encountered, $i.e.$, with the structural response at $k_h$, causing a large amplification of the heave LCO and some amplification of the pitch LCO.

Finally, with freeplay $\alpha_s = 0.5^\circ$ resonance is encountered for all heave frequency ratios. For frequency ratios $\hat{k}_h < 0.49$, the aerodynamic forcing locks in to the heave natural frequency superharmonic $3k_h$, and a 3:1 resonance is encountered with the structure oscillating oscillating at $k_h$. At the lowest frequency ratio $\hat{k}_h = 0.4$ the LCO amplitude drastically increases and appears to be at the maximum amplification of the 3:1 lock-in range, reducing as the frequency ratio increases. Presumably, the heave LCO amplitude would reduce as the frequency ratio reduces towards $\hat{k}_h = 0.33$ which is the onset of the 3:1 lock-in region. This level of amplification could also be explained by the low structural stiffness required to achieve the lower frequency ratios, and that the system approaches linear flutter as the frequency ratio reduces. For this range, $\hat{k}_h < 0.49$, the pitch LCO is approximately double the amplitude of the case without freeplay, or that with freeplay $\alpha_s = 0.25^\circ$, and does not see noteworthy amplification as a function of frequency ratio. This suggests that the the increase in amplitude observed with respect to the other two freeplay angles is an artifact of the freeplay itself, rather than lock-in or resonance behavior. At the frequency ratio $\hat{k}_h = 0.49$ there is an abrupt transition where the lock-in drops to the lower harmonic $2k_h$ with 2:1 resonance. It is interesting that the heave LCO amplitude continues to decay in this region, up to {\color{black}$\hat{k}_h = 0.52$}, rather than consistently increase as was observed with freeplay $\alpha_s = 0.25^\circ$. Then, for {\color{black}$\hat{k}_h > 0.52$} the heave LCO begins to grow in amplitude. The pitch LCO grows for frequency ratios $\hat{k}_h > 0.5$ - similar to what is observed with freeplay $\alpha_s = 0.25^\circ$. 

\subsection{2:1 Lock-In Mechanism}
\begin{figure}[h!]
    \centering
        \includegraphics[width=1\textwidth]{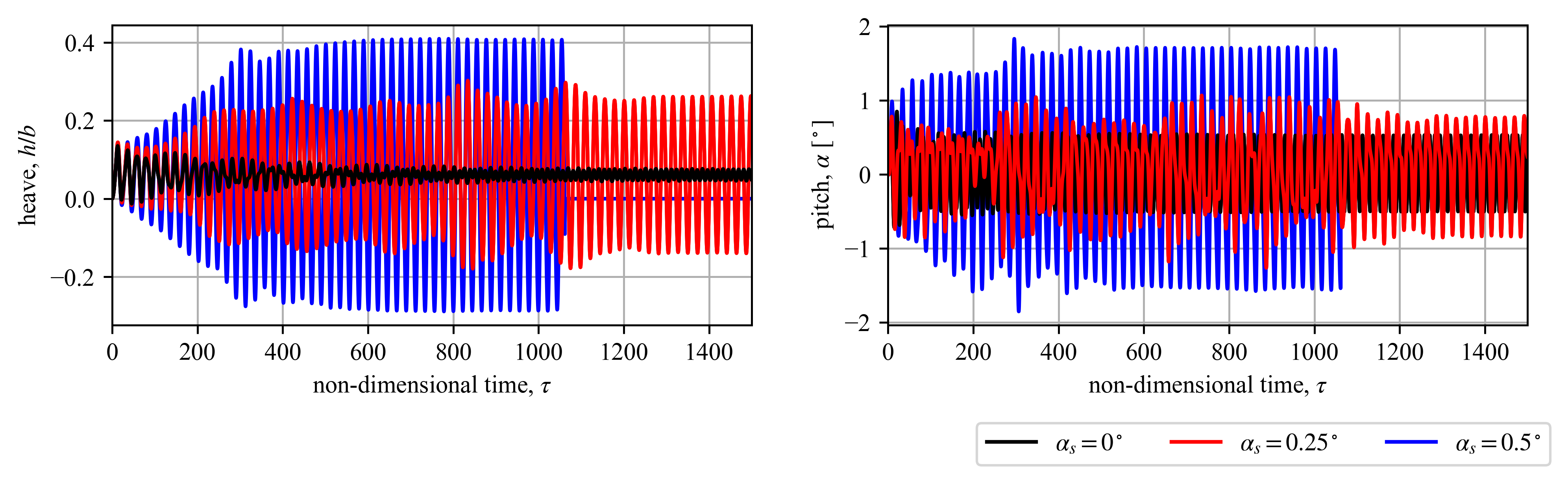}
    \caption{Time responses with $\mu = 75$, $\hat{k}_h = 0.555$, $\hat{k}_\alpha = 0.755$}
    \label{fig:time_21}
\end{figure}

\begin{figure}[h!]
    \centering
        \includegraphics[width=1\textwidth]{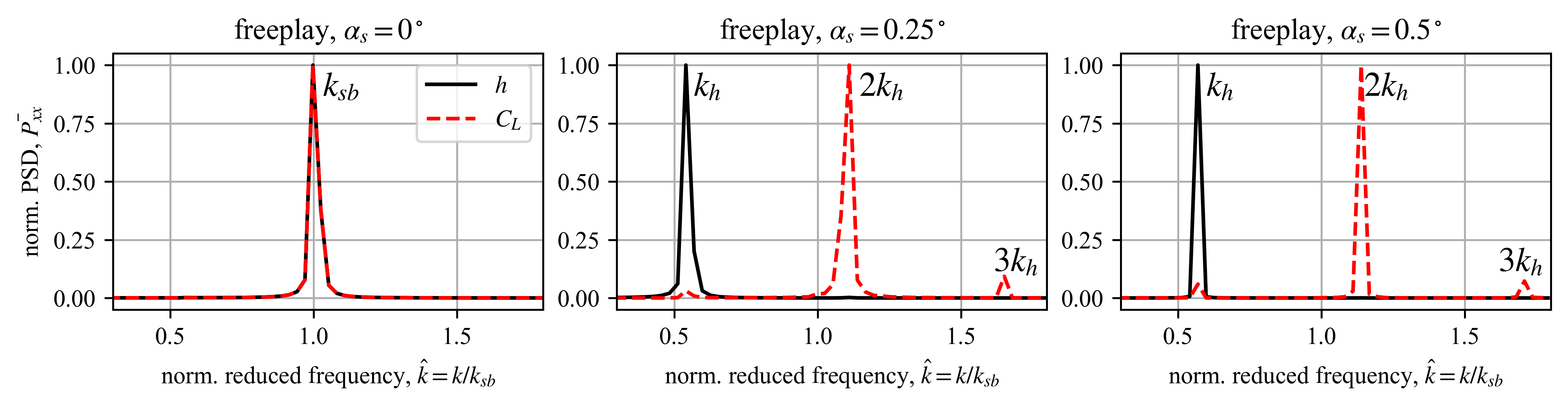}
    \caption{Power spectral density of the stable LCO with $\mu = 75$, $\hat{k}_h = 0.555$, $\hat{k}_\alpha = 0.755$}
    \label{fig:freq_21}
\end{figure}

The heave and pitch responses without and with varying freeplay angles are presented in the time domain in Fig.~\ref{fig:time_21}, and the power spectral density (PSD) of the heave and lift LCO are presented in Fig.~\ref{fig:freq_21}. The PSD is normalized to 1 (by the respective maximum power) for visualization purposes. The response without freeplay is characterized by a single-period LCO that is oscillating at the buffet frequency - characteristic of a forced harmonic oscillator. This is depicted in Fig.~\ref{fig:freq_21} where both the lift and heave responses are dominated by a single harmonic at $k_{sb}$. 

\begin{figure}[h!]
    \centering
        \includegraphics[width=1\textwidth]{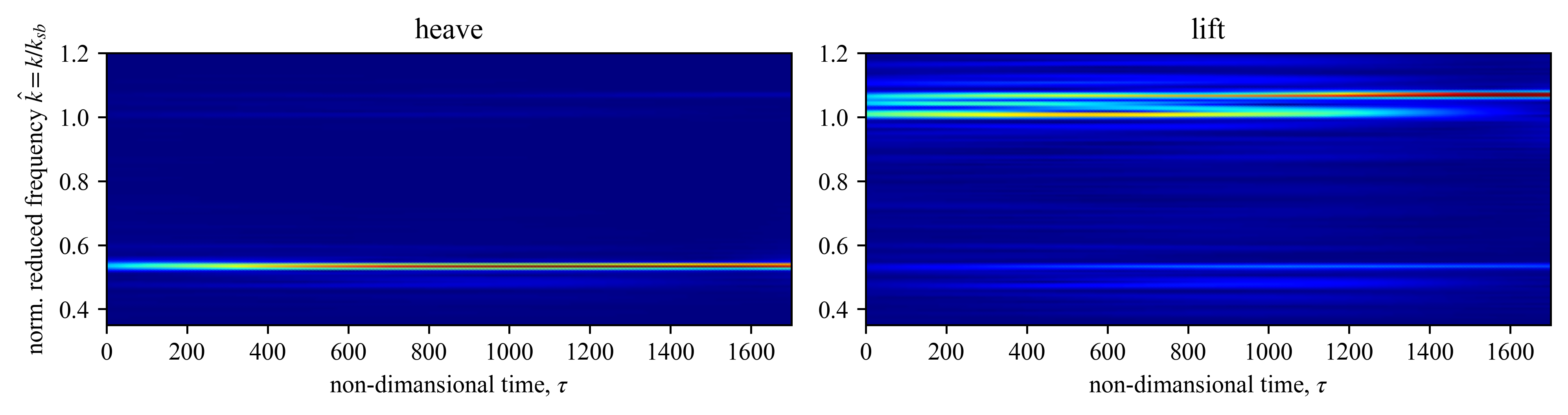}
    \caption{Short-Time Fourier Transform with $\mu = 75$, $\hat{k}_h = 0.555$, $\hat{k}_\alpha = 0.755$ and $\alpha_s = 0.25^\circ$}
    \label{fig:time_freq_21}
\end{figure}

The responses with freeplay $\alpha_s = 0.25^\circ$ is characterized by significant amplification of the LCO in the heave mode ($16\times$ relative to the case without freeplay), and moderate amplification in the pitch mode ($1.75\times$ relative to the case without freeplay). The initial portion of the response ($\tau<100$) is characterized by decay of the heave response and period-two oscillations in the pitch response. Beyond this, the heave response begins to grow and at $\tau \approx 300$ the pitch response abruptly transitions to larger amplitude singe-period oscillations. From this point up to $\tau \approx 1300$ a transient can be observed in the time responses with - characterized by a beating phenomenon. Observation of the Short-Time Fourier Transform in Fig.~\ref{fig:time_freq_21} indicates that this transient is characterized by the modular frequencies $k_{sb}$ and $2k_h$, which explains the beating pattern through this region. Finally, a stable LCO is encountered and inspection of Fig.~\ref{fig:freq_21} indicates that the lift forces exhibit lock-in to the the heave mode superharmonic $2k_h$, with $k_h$ and $3k_h$ also visible. The heave oscillations are at the heave natural frequency, characteristic of a subharmonic resonance, $i.e.$, where the structure responds at a subharmonic of the forcing frequency. Although not shown, the frequencies in the pitching moment LCO are consistent with the lift and the pitch rotation LCO consistent with heave. 

With freeplay $\alpha_s = 0.5^\circ$ the behavior is consistent albeit more aggressive, with the heave mode rapidly growing to a stable LCO that is $24\times$ larger than the case without freeplay. The pitch mode amplification relative to the case without freeplay is $3\times$. The lock-in and resonance characteristics are consistent with those that were described for $\alpha_s = 0.25^\circ$ as can be observed in Fig.~\ref{fig:freq_21}. As a final note, in previous work~\cite{candon25d} it was shown that with freeplay $\alpha_s = 0.1^\circ$, no form of lock-in or resonance is encountered.

\begin{figure}[h!]
    \centering
        \includegraphics[width=1\textwidth]{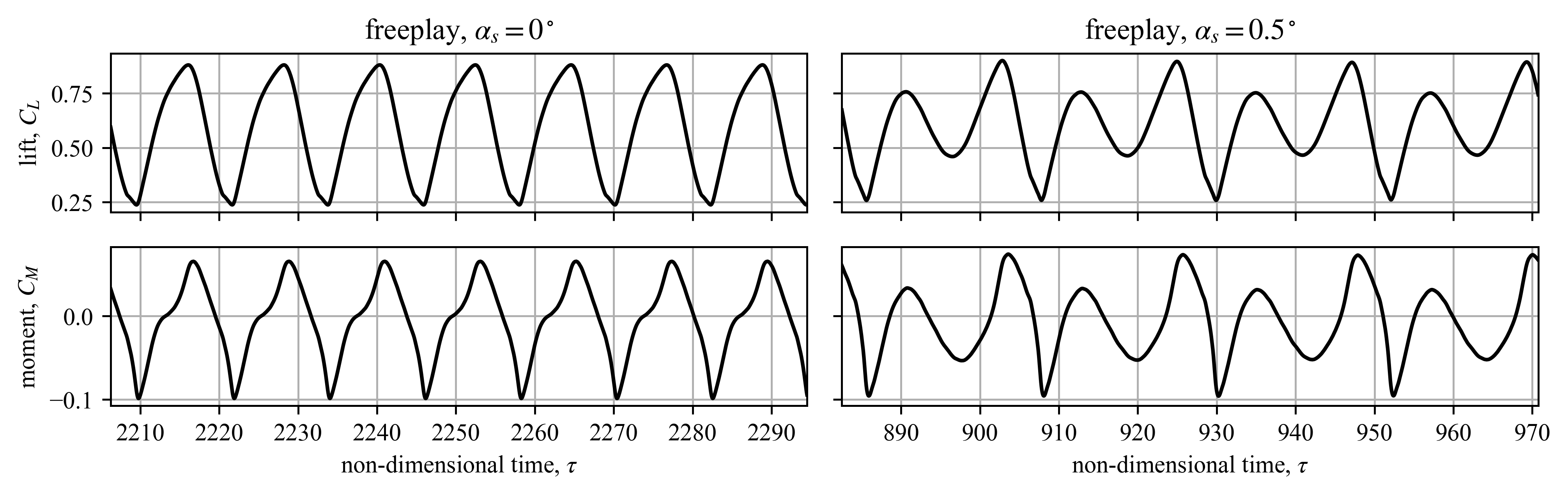}
    \caption{Aerodynamic forces and moments with $\mu = 75$, $\hat{k}_h = 0.555$ and $\hat{k}_\alpha = 0.755$}
    \label{fig:aero_21}
\end{figure}

To gain a deeper understanding of the lock-in mechanism, the aerodynamic forces and moments are now assessed, along with the time accurate shock location. The shock location is extracted from the pressure coefficient on the airfoil surface, which is recorded every 20 physical time-steps. Figure~\ref{fig:aero_21} presents a portion of the force and moment responses, after a stable LCO is encountered, without freeplay and with freeplay $\alpha_s = 0.5^\circ$. The case without freeplay is characterized by single-period oscillations with a strong nonlinear form - not significantly different to the lift and moment oscillations on the rigid airfoil. The aerodynamics with freeplay $\alpha_s = 0.5^\circ$ is vastly different, demonstrating period-two oscillations of the forces and moments with a rich nonlinear form - driving the 2:1 lock-in mechanism. No notable amplification of the lift and mild amplification of the moment can be observed.

\begin{figure}[h!]
    \centering
        \includegraphics[width=1\textwidth]{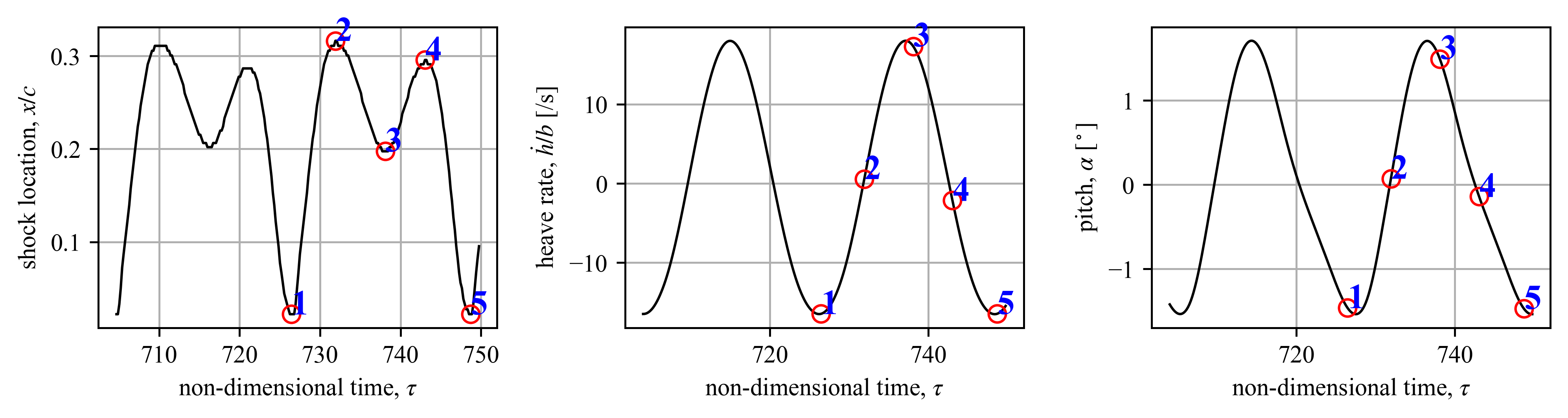}
    \caption{Time responses depicting shock location for one lock-in cycle with $\mu = 75$, $\hat{k}_h = 0.555$, $\hat{k}_\alpha = 0.755$ and $\alpha_s = 0.5^\circ$}
    \label{fig:time_shock_f055_fp05}
\end{figure}

\begin{figure}[h!]
    \centering
        \includegraphics[width=1\textwidth]{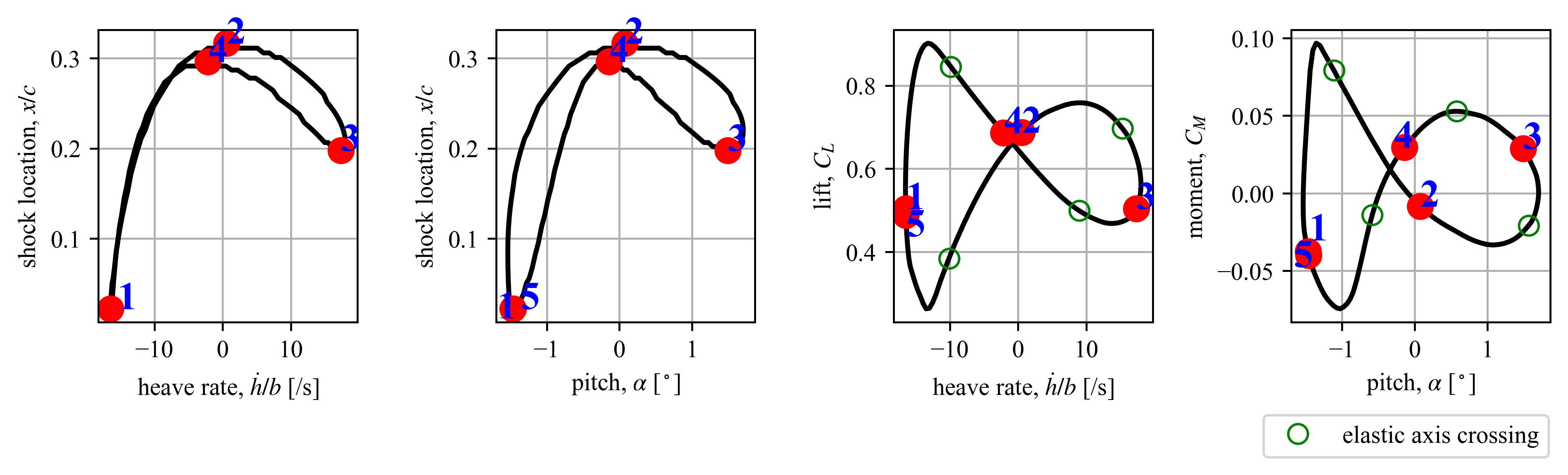}
    \caption{Lissajous curves depicting shock location for one lock-in cycle with  $\mu = 75$, $\hat{k}_h = 0.555$, $\hat{k}_\alpha = 0.755$ and $\alpha_s = 0.5^\circ$}
    \label{fig:liss_shock_f055_fp05}
\end{figure}

Figure~\ref{fig:time_shock_f055_fp05} presents the time-accurate shock location with the extrema (shock locations at the turning points) marked across one cycle. Those locations are then plotted over the time accurate heave rate and pitch. The time accurate shock location indicates period-two oscillations which aligns wirg the period-two lift and moment oscillations observed in Fig.~\ref{fig:aero_21}. The shock location and structural motion is synchronized which can be observed at points 1, 3 and 5 where the extrema of the shock location, the heave rate and pitch rotation are aligned. In Fig.~\ref{fig:liss_shock_f055_fp05} it is shown that the phase angle between shock location and heave rate, and shock location and pitch rotation, is small - computed to be $8.6^\circ$ and $15.8^\circ$, respectively. The Lissajous curves depicting lift as a function of heave rate, and moment as a function of pitch rotation, are both characterized by two closed sections as a result of the 2:1 relationship between the forcing frequency at $2k_h$ and the pitch response at $k_h$. The pitch-moment curve demonstrates that pitch cycle moves through its trough (points 1-2), it coincides with aggressive downstream shock motion and a vertical increase in the pitching moment as the shock approaches the elastic axis, after which the airfoil begins pitching up. The relationship between heave rate and the lift is qualitatively similar with the trajectory direction reversed. Specifically, between points 4 and 5 where aggressive upstream shock motion is observed, the shock crosses the elastic axis after which an aggressive drop in lift occurs. As the shock reaches its farthest upstream location and reverses, the lift continues to drop, then reverses rapidly as the shock approaches the elastic axis.

\subsection{3:1 Lock-In Mechanism}

\begin{figure}[h!]
    \centering
        \includegraphics[width=1\textwidth]{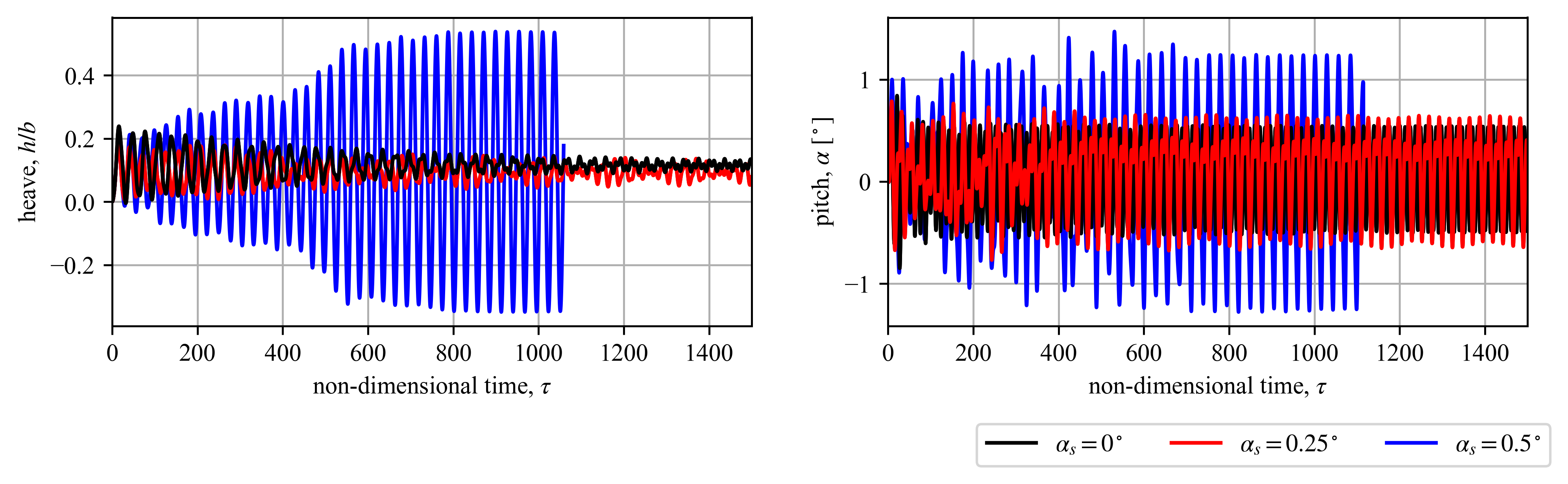}
    \caption{Time responses with $\mu = 75$, $\hat{k}_h = 0.45$, $\hat{k}_\alpha = 0.755$}
    \label{fig:time_31}
\end{figure}

\begin{figure}[h!]
    \centering
        \includegraphics[width=1\textwidth]{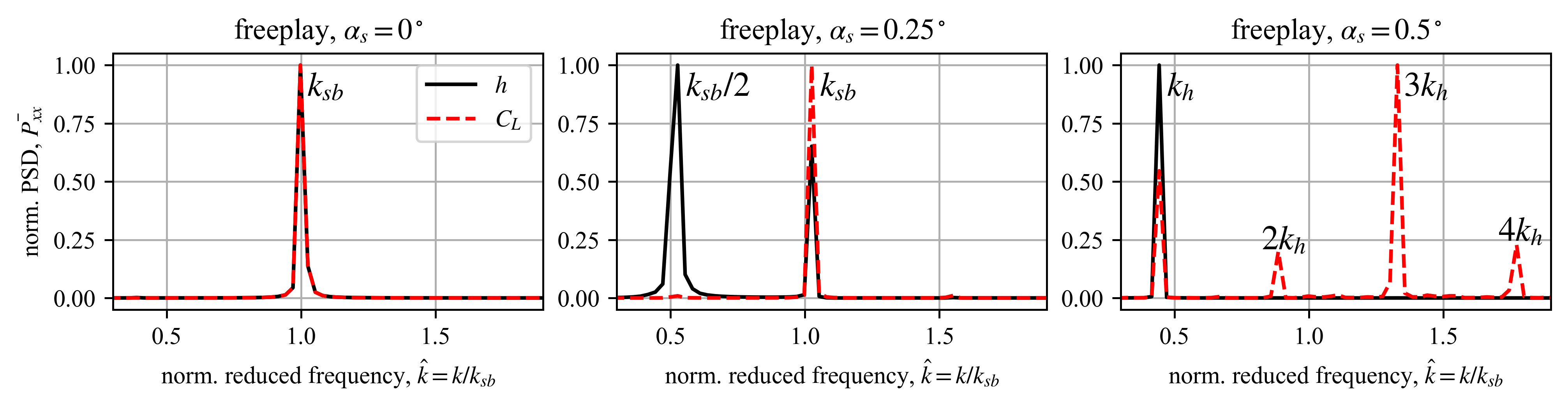}
    \caption{Power spectral density of the stable LCO with $\mu = 75$, $\hat{k}_h = 0.45$, $\hat{k}_\alpha = 0.755$}
    \label{fig:freq_31}
\end{figure}

 The 3:1 lock-in mechanism is now investigated, with the time responses presented in Fig.~\ref{fig:time_31}. The cases without freeplay and with freeplay $\alpha_s = 0.25^\circ$ are truncated for visualization purposes, however, were run for a total non-dimensional time of 4000 to achieve a stable LCO. The maximum-normalized PSD of the lift and heave LCOs are given in Fig.~\ref{fig:freq_31}. Again, the pitch and moment PSD is not shown as they are equivalent. It can be seen that without freeplay, the response is consistent with that described previously, $i.e.$, representative of a forced harmonic oscillator. 
 
 The response with freeplay $\alpha_s = 0.25^\circ$ does not encounter any form of resonance; the forcing does not lock-in to a structural superharmonic and is dominated by the shock buffet frequency or, more precisely, a frequency that is slightly offset from $k_{sb}$ as was described in Section~\ref{sec:frat_sweep}. In contrast to the case without freeplay, the structural response is at $k_{sb}/2$. This shows that resonance does not occur because the structural natural frequency $k_h < 0.5k_{sb}$ and similarly lock-in cannot occur given that the structural superharmonic $2k_h < k_{sb}$. The system does not exhibit lock-in to $3k_h$, presumably because this higher harmonic has insufficient vibrational energy with this amount of freeplay. The LCO amplitude does not encounter any notable amplification in either mode. 
 
 With freeplay $\alpha_s = 0.5^\circ$ the aerodynamic forces exhibit lock-in to the heave mode superharmonic $3k_h$ while the structure responds at $k_h$, characteristic of a subharmonic resonance with the structure responding at a subharmonic of the forcing frequency. The amplification of the LCO relative to the case without freeplay is $\sim3700\%$ in the heave mode and $\sim100\%$ in the pitch mode.

\begin{figure}[h!]
    \centering
        \includegraphics[width=1\textwidth]{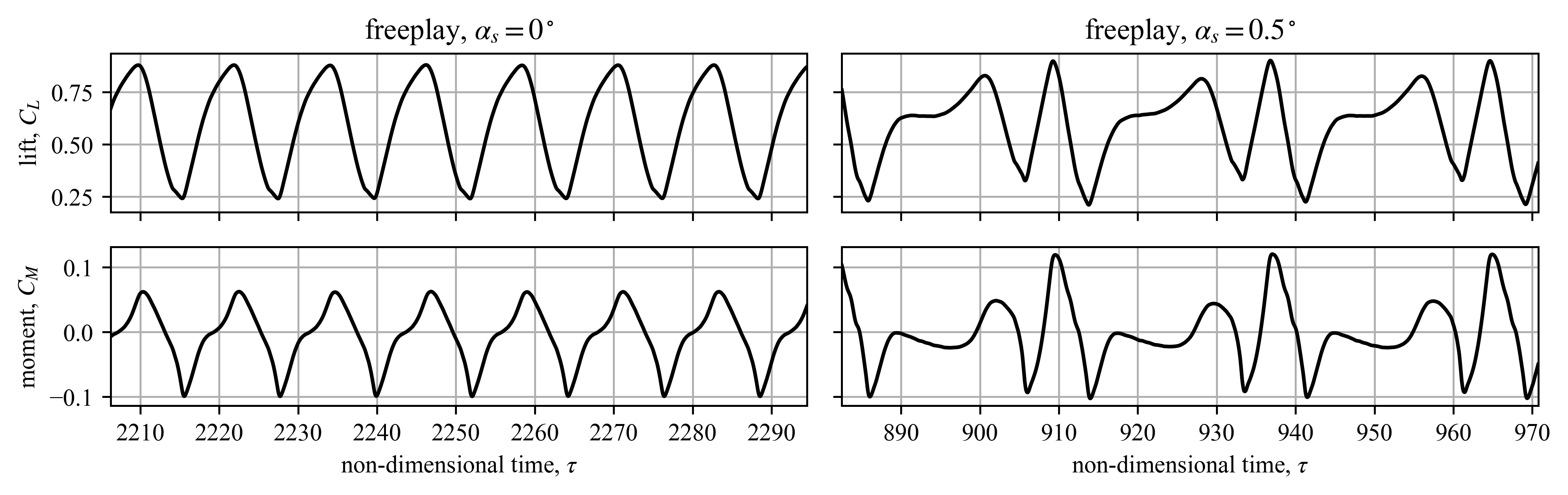}
    \caption{Aerodynamic forces and moments with $\mu = 75$, $\hat{k}_h = 0.45$ and $\hat{k}_\alpha = 0.755$}
    \label{fig:aero_31}
\end{figure}

\begin{figure}[h!]
    \centering
        \includegraphics[width=1\textwidth]{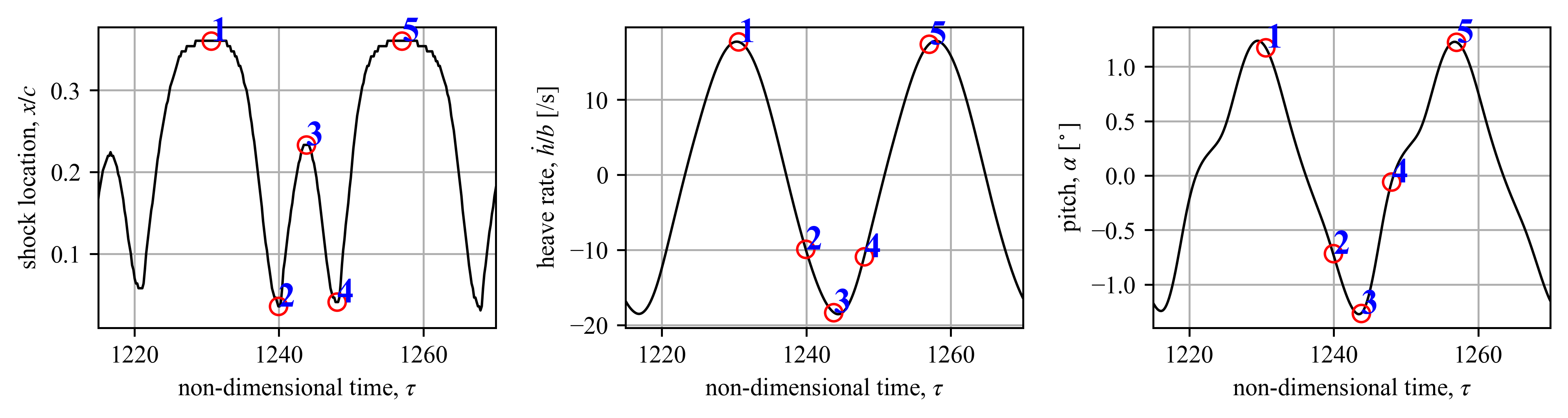}
    \caption{Time responses depicting shock location for one lock-in cycle with $\mu = 75$, $\hat{k}_h = 0.45$, $\hat{k}_\alpha = 0.755$ and $\alpha_s = 0.5^\circ$}
    \label{fig:time_shock_f045_fp05}
\end{figure}

Figure~\ref{fig:aero_31} presents the time-varying lift and moment for the case exhibiting 3:1 lock-in, depicting a period-three limit cycle with a rich and complex nonlinear form. Significant amplification of the pitching moment is observed. This is in contrast to the case without freeplay characterized by single-period oscillations and the period-two oscillations that were shown in Fig.~\ref{fig:aero_21}. Figure~\ref{fig:time_shock_f045_fp05} presents the shock location, with the turning points marked and plotted over the heave rate and pitch rotation. Notably, the shock location for this case is characterized by a flat region, $i.e.$, for a portion of the aeroelastic cycle, the shock remains almost static in the downstream location which appears to coincide with the peak of the heave rate cycle. Another notable feature is the period-two nature of the shock cycle, which is surprising given that the aerodynamic response is period-three. Similar to the 2:1 lock-in case (Fig.~\ref{fig:time_shock_f055_fp05}) there is alignment of the extrema of the heave rate and pitch angle, and the shock turning points. That said, for this case, the maximum heave rate/pitch angle coincide with the maximum downstream shock location, while the opposite was observed for the 2:1 lock-in case.  

\begin{figure}[h!]
    \centering
        \includegraphics[width=0.66\textwidth]{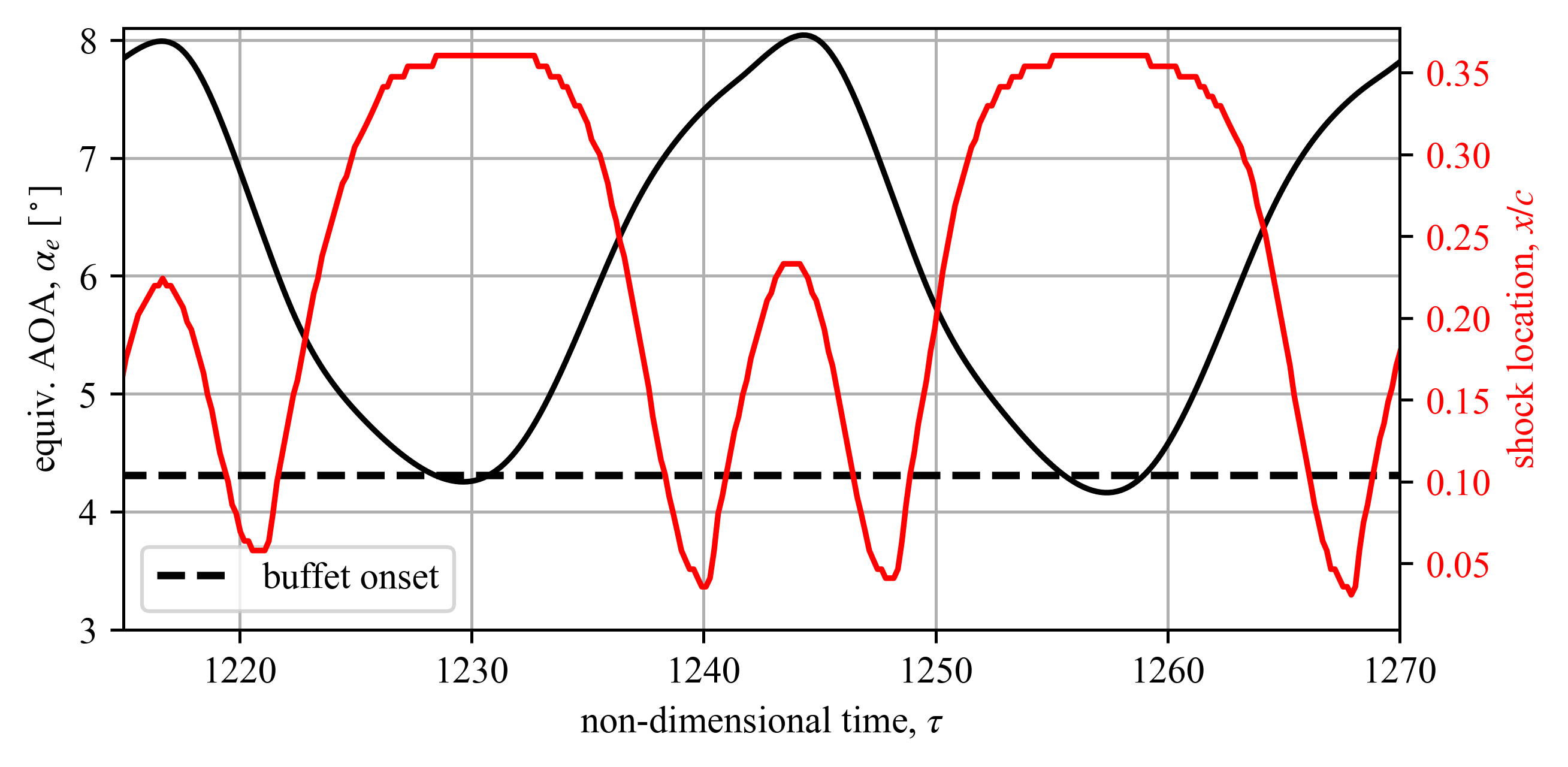}
    \caption{Equivalent angle-of-attack and shock location with $\mu = 75$, $\hat{k}_h = 0.45$, $\hat{k}_\alpha = 0.755$ and $\alpha_s = 0.5^\circ$}
    \label{fig:eaoa_f045_fp05}
\end{figure}

The portion of the cycle for which the shock is stationary is most interesting and is investigated by assessing the equivalent AOA, $\alpha_e = \alpha_0 + \alpha - \dot{h}/U_\infty$, which takes into account the mean AOA, $\alpha_0 = 6^\circ$, the time-varying pitch angle, $\alpha$, and the induced AOA due to the heave rate. By overlaying the equivalent AOA and the shock location, as presented in Fig.~\ref{fig:eaoa_f045_fp05}, it can be seen that for the portion of the cycle for which the shock is stationary, the large heave rate causes the equivalent AOA to drop below the buffet onset angle $\alpha_{sb} = 4.31^\circ$. This temporarily pinned shock location at $x/c = 0.36$ seems to agree reasonably well with the steady-state shock location presented in Fig.~\ref{fig:mesh_independenceA}, which is at an AOA of $\alpha_0 = 4^\circ$.  
\clearpage

\begin{figure}[h!]
    \centering
        \includegraphics[width=1\textwidth]{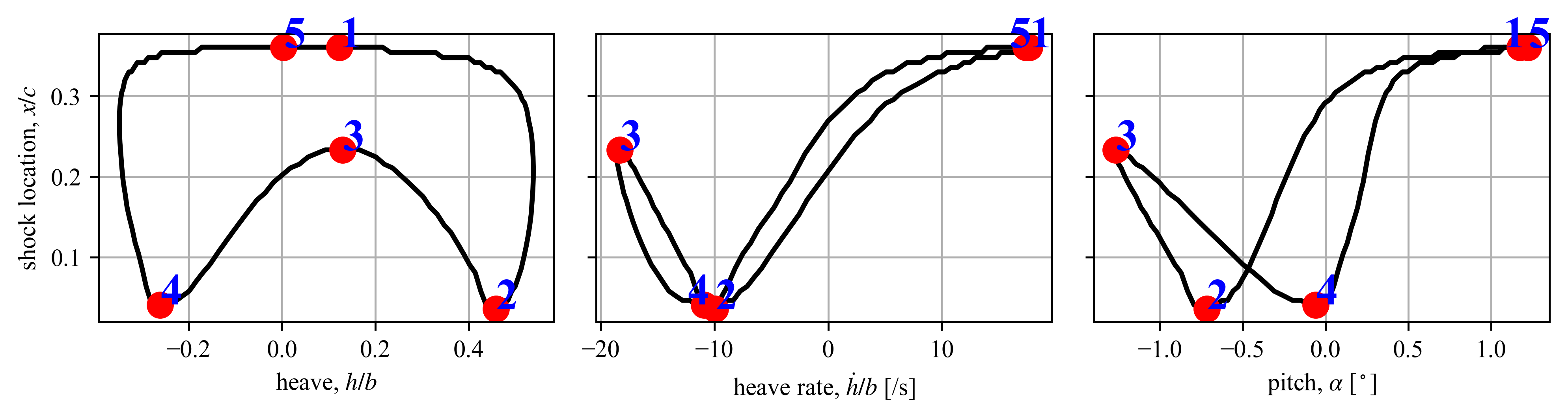}
    \caption{Shock location as a function of structural response for $\mu = 75$, $\hat{k}_h = 0.45$, $\hat{k}_\alpha = 0.755$ and $\alpha_s = 0.5^\circ$}
    \label{fig:liss_shock_a_f045_fp05}
\end{figure}

\begin{figure}[h!]
    \centering
        \includegraphics[width=0.66\textwidth]{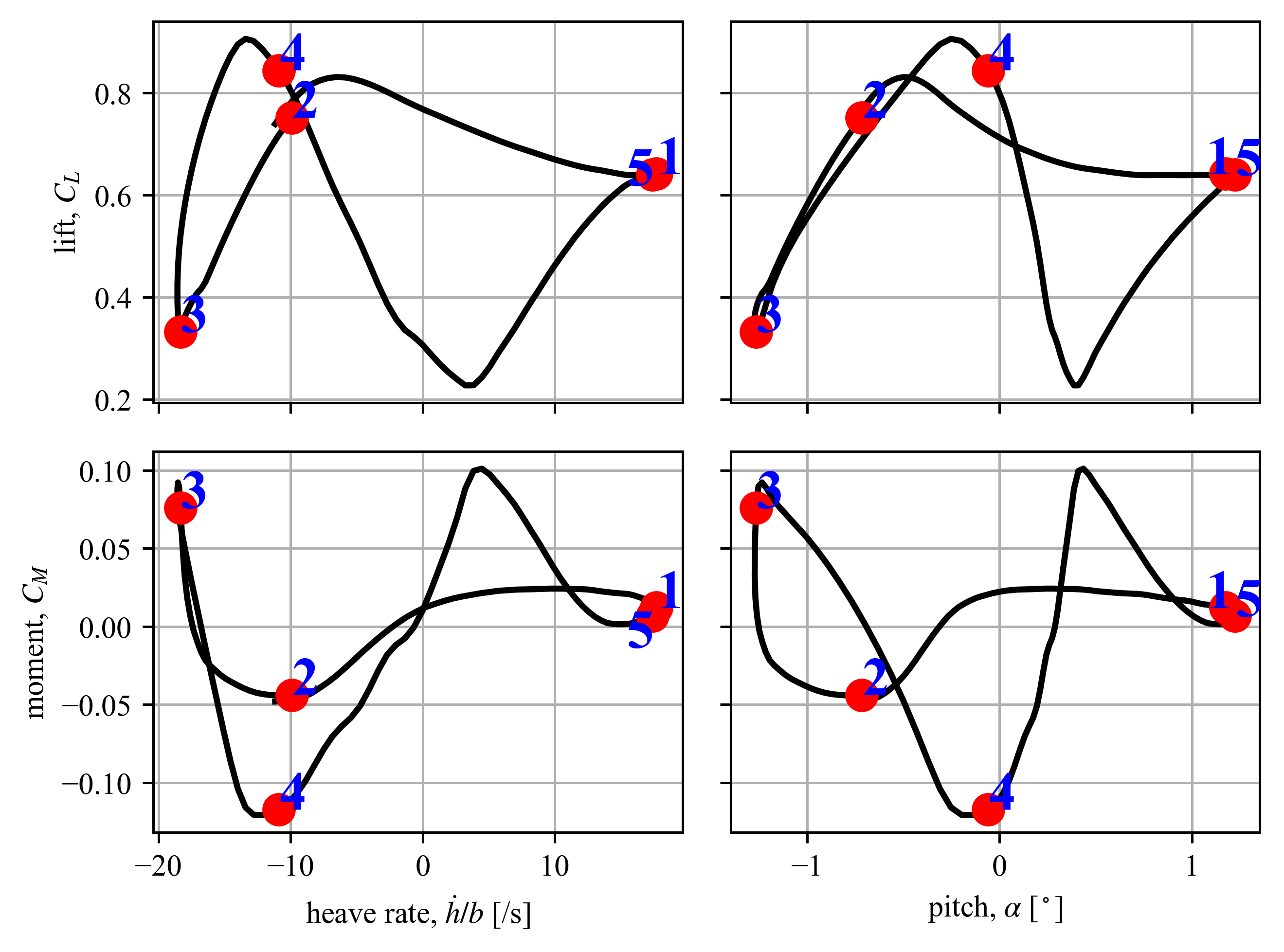}
    \caption{Lissajous curves for $\mu = 75$, $\hat{k}_h = 0.45$, $\hat{k}_\alpha = 0.755$ and $\alpha_s = 0.5^\circ$}
    \label{fig:liss_shock_b_f045_fp05}
\end{figure}

Figure~\ref{fig:liss_shock_a_f045_fp05} presents the shock location as a function of structural motion. The relationship with the heave displacement shows that as the airfoil traverses from the minimum to maximum heave positions, which also coincides with the maximum heave velocity, the shock encounters the stationary portion of the cycle. The phase angle between the shock location and heave rate is small, and somewhat larger with respect to pitch angle. Both depict 2:1 frequency content. A selection of Lissajous curves for the system are presented in Fig.~\ref{fig:liss_shock_b_f045_fp05}. Firstly, it is noteworthy that curves for lift depict dominance of 2:1 frequency content (with 3:1 content also visible in relation to the pitch angle, while the the moments clearly demonstrate higher harmonic relations with 3:1 and even 4:1 frequency content. Of course, it is expected that the moments will contain a richer spectrum of the nonlinear dynamics given that they are being about the quarter-chord location (close to the aerodynamic center), and that the shock is crossing the moment summation point. The influence of the stationary shock is intuitively more visible for the moments, which remain relatively constant for a portion of the curve between points 1 and 2, and as it approaches point 5, $i.e.$, indicating some lag between the shock location and the corresponding change in moment. The influence of the stationary shock is less visible in the lift curves, which are influenced more by the airfoil motion. The relationship between heave rate and moment depicts sharp gradients between points 2 and 3, and 3 and 4, where an aggressive sub-oscillation of the shock from its maximum upstream location coincides with the maximum equivalent AOA. This can also be seen between points 2 and 3 in the relation between pitch angle and moment, and between points 3 and 4 in the lift curves. 

\clearpage


\subsection{Sensitivity to Mass Ratio and Structural Damping}

\begin{figure}[h!]
    \centering
        \includegraphics[width=1\textwidth]{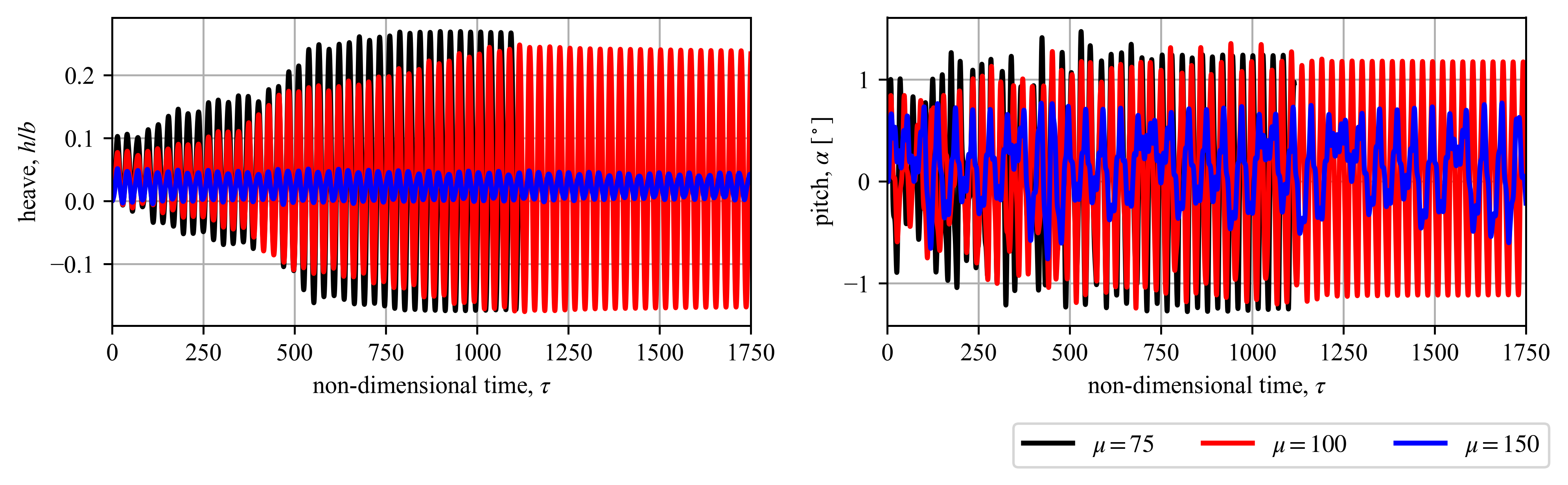}
    \caption{Time responses at $\hat{k}_h = 0.45$, $\hat{k}_\alpha = 0.755$ and $\alpha_s = 0.5^\circ$ with various mass ratios}
    \label{fig:mu_045_resp}
\end{figure}

\begin{figure}[h!]
    \centering
        \includegraphics[width=1\textwidth]{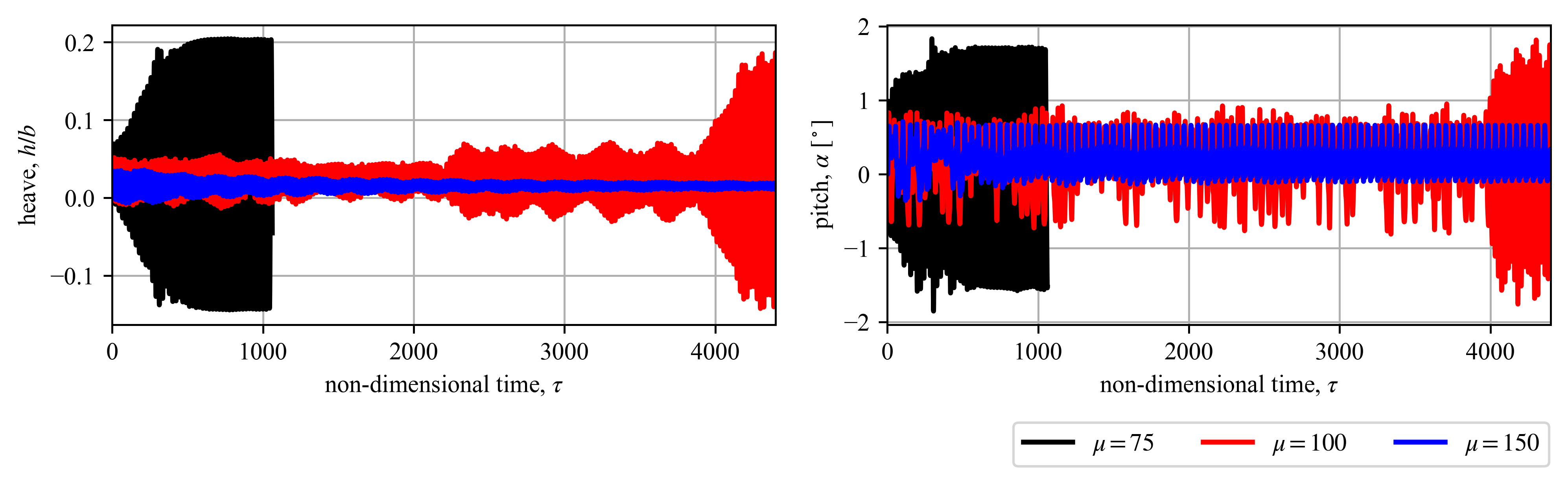}
    \caption{Time responses at $\hat{k}_h = 0.555$, $\hat{k}_\alpha = 0.755$ and $\alpha_s = 0.5^\circ$ with various mass ratios}
    \label{fig:mu_055}
\end{figure}

The final analysis is to assess the influence of mass ratio and structural damping on the lock-in mechanism, noting that previous authors~\cite{raveh14, giannelis16} have shown that lock-in, or rather LCO amplification due to lock-in, is relatively insensitive to mass ratio and sensitive to structural damping when the structural model is linear. Figure~\ref{fig:mu_045_resp} presents the influence of the mass ratio on the 3:1 lock-in mechanism. It can be seen that an increase in the mass ratio of 33\% slows the growth of the LCO and marginally reduces the amplitude, while by doubling the mass ratio, lock-in does not occur and the resonance is completely suppressed. Similar findings can be observed for the 2:1 lock-in mechanism with $\hat{k}_h=0.555$ as is presented in Fig.~\ref{fig:mu_055}, although lock-in is significantly delayed with a mass ratio $\mu = 100$. For this case, rather than the delayed lock-in being due to very slow growth, the large amplitude LCO appears abruptly and grows rapidly. Prior to lock-in, the dynamics of this case appear chaotic.

\begin{figure}[h!]
    \centering
        \includegraphics[width=1\textwidth]{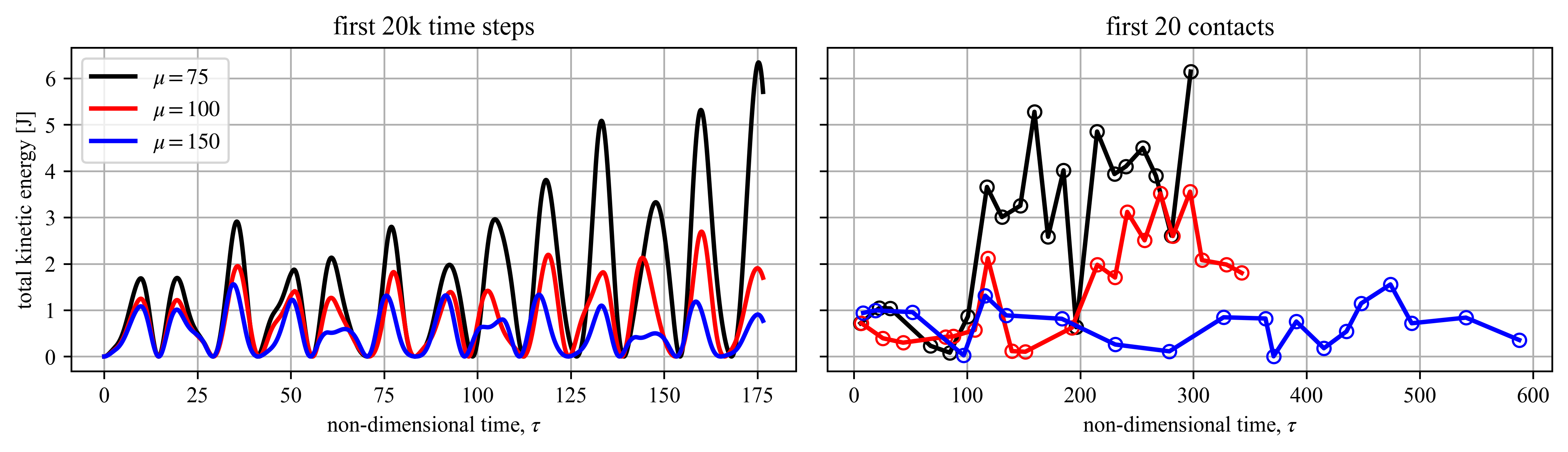}
    \caption{Total kinetic energy at $\hat{k}_h = 0.45$, $\hat{k}_\alpha = 0.755$ and $\alpha_s = 0.5^\circ$ with various mass ratios}
    \label{fig:mu_045_tke}
\end{figure}

These results may seem surprising, given the the findings of Giannelis \textit{et al.}~\cite{giannelis16}; who demonstrated insensitivity with an up to eight-fold increase of the mass for a linear structural model. However, a physical explanation can be provided. The nonlinear lock-in and resonance mechanisms herein require that the higher harmonics that are activated when the airfoil impacts the freeplay dead-zone boundary are of sufficient energy for the aerodynamic forces to lock-in. The additional mass in the system reduces the kinetic energy (KE) with which the impact occurs, as can be observed in Fig.~\ref{fig:mu_045_tke}. Notably, the first few impacts ($\tau<100$) occur with a similar KE, then the first impact after $\tau=100$ is with a nearly 4$\times$ increase in KE (with respect to the previous maximum) for $\mu = 75$, and a 2$\times$ increase for $\mu = 100$, while no notable increase can be observed for $\mu = 150$. Moreover, the number of contacts per time period notably reduces for the $\mu = 150$ case, predominantly due to an inconsistent impact of the lower dead-zone boundary, further contributing to the insufficient superharmonic energy content. 

\clearpage

\begin{figure}[h!]
    \centering
        \includegraphics[width=1\textwidth]{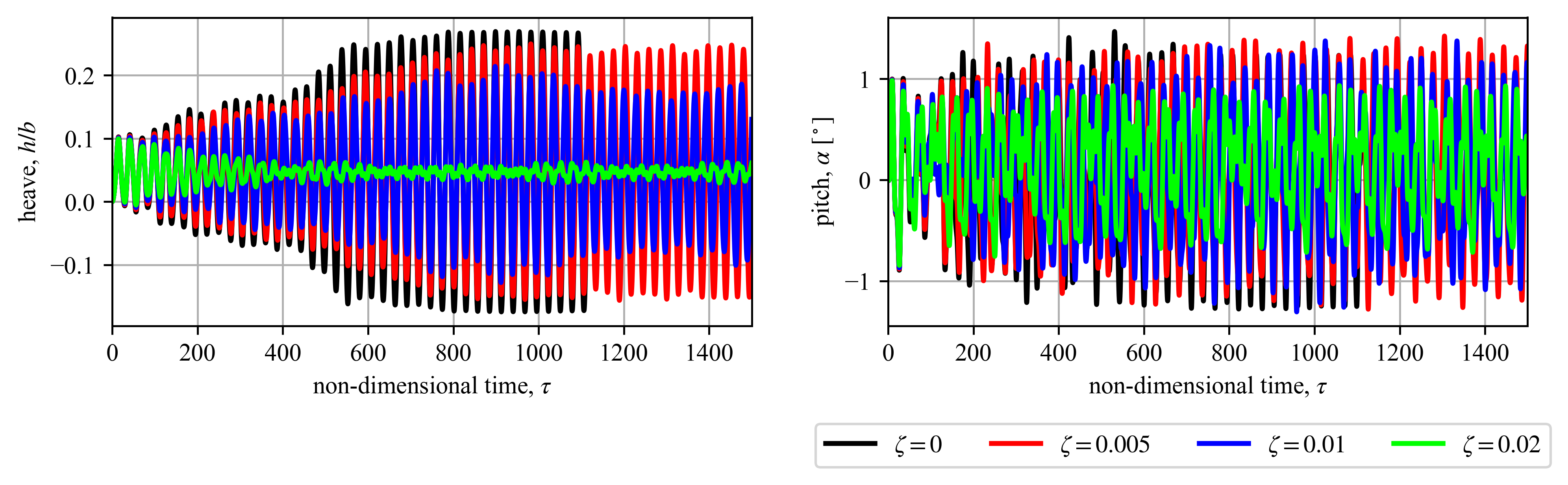}
    \caption{Time responses at $\mu = 75$, $\hat{k}_h = 0.45$, $\hat{k}_\alpha = 0.755$ and $\alpha_s = 0.5^\circ$ with structural damping}
    \label{fig:damp_045}
\end{figure}

The influence of structural damping on the 3:1 lock-in mechanism is presented in Fig.~\ref{fig:damp_045}. Equal structural damping is applied to both modes. It can be seen that with as little as 0.5\% structural damping, the LCO amplitude is considerably reduced, and with 2\% structural damping (which is the typically assumed value when the exact structural damping is unknown) the lock-in and resonance mechanisms are completely suppressed. This high sensitivity to structural damping aligns with the findings of Raveh and Dowell~\cite{raveh14} and Giannelis \textit{et al.}~\cite{giannelis16}.


\section{Conclusions}

Transonic shock buffet is a complex unsteady aerodynamic phenomenon that plagues modern aircraft, characterized by large amplitude self-sustaining shock oscillations. Despite numerical investigations of the aeroelastic response to shock buffet oscillations receiving signification interest in the last decade, the influence structural nonlinearities have not yet been considered. This paper presents a numerical investigation of the aeroelastic interaction between shock buffet and freeplay-induced nonlinear structural dynamics, where Unsteady Reynolds-Averaged Navier Stokes simulations of the NACA0012 airfoil are coupled with the nonlinear 2-DOF pitch-heave equations-of-motion. 

It is shown that the presence of freeplay can induce nonlinear resonance and lock-in behavior that does not exist in the absence of freeplay ($i.e.$, when the structural model is linear). High-amplitude LCOs are shown to occur when the ratio of the structural natural frequency to the shock buffet frequency is in the vicinity of 0.4 - 0.75, which is much lower than has been shown previously for a linear structural model. The nonlinear dynamical mechanism by which the aerodynamic forces exhibit lock-in is through the heave natural frequency superharmonics, leading to 2:1 and 3:1 resonances mechanisms. More specifically, the presence of freeplay leads to the redistribution of vibrational energy to superharmonics of the heave natural frequency, to which the aerodynamic forces exhibit lock-in. The onset of the 2:1 lock-in mechanism (lock-in to the superharmonic of order 2) occurs with relatively small freeplay quantities ($0.1^\circ < \alpha_s < 0.25^\circ$). To encounter lock-in to higher harmonics (above 2) a larger freeplay angle is needed to achieve sufficient vibrational energy, with onset in the range $0.25^\circ < \alpha_s < 0.5^\circ$. It is shown that with a sufficiently small freeplay ($\alpha_s\leq0.1^\circ$ in the cases presented in this paper) the nonlinear resonance does not occur and the system behaves as a forced harmonic oscillator. Various rich nonlinear dynamical features of the lock-in mechanism are unveiled and studied in detail, including the analysis of Lissajous curves, the time-accurate shock location and time-frequency analysis. The lock-in and resonance mechanisms are shown to be highly sensitive to both the structural-to-fluid mass ratio and structural damping. 

This work is a part of a larger research program on the study of nonlinear aeroelastic interactions. Having identified an interesting and potentially catastrophic phenomenon, exciting new opportunities exist to further this work. In particular, the authors recommend experimental verification of the observed phenomena, studying the buffet-freeplay dynamics in real-world aircraft configurations, and understanding the interactions with flap freeplay. 

\section*{Acknowledgments}
This work would not have been possible without the financial support provided by the Asian Office of Aerospace Research and Development (AOARD) and Air Force Office of Scientific Research (AFOSR) for project FA2386-24-1-4044: Data-Driven Reduced Order Modelling and Preliminary Experimentation for Combined Transonic Buffet and Freeplay Induced Limit Cycle Oscillations. We are grateful for the partial financial support provided by the Australian Defence Science and Technology Group (DSTG). The computational resources provided by ANSYS, and the support of Dr Valerio Viti and Dr Luke Munholand, are also greatly appreciated. The contributions of our partners on the project, Prof Sergio Ricci from Politecnico di Milano and Prof Hideaki Ogawa from Kyushu University have been instrumental to this work. Finally, we are deeply grateful for the ongoing advice and support of Prof Earl Dowell and Prof Lawrie Virgin from Duke University, Dr Walter Silva from NASA Langley Research Center, and the efforts and support of Dr Pawel Chwalowski, Dr Bret Stanford, Dr Brent Pomeroy, and all others in the DPW-8/AePW-4 organizing committee.  

\bibliographystyle{elsarticle-num} 
\bibliography{gensys_doi}
        

\end{document}